%% file: lecture_notes.tex
\documentclass[12pt]{article}
\usepackage[a4paper, left=2.25cm, top=3cm, bottom=3cm, right=2.25cm]{geometry}
\usepackage{latexsym}
\usepackage{amssymb}
\usepackage{amsmath}
\usepackage{amstext}
\usepackage{bm}
\usepackage{fixmath}
\usepackage{makeidx}
\usepackage{graphicx}
\usepackage{rotating}
\usepackage{url}
\usepackage{times}
\usepackage{listings}
\usepackage{xcolor}
\usepackage{qcircuit}
\usepackage{multirow}
\usepackage{fancyhdr}
\usepackage[colorlinks=true,linkcolor=blue,citecolor=red,urlcolor=blue,linktocpage=true]{hyperref}
\usepackage{authblk}
\usepackage[style=numeric,maxbibnames=10,alldates=short,giveninits=true,backref=false,hyperref=true,sorting=none,
url=false,doi=false,isbn=false,eprint=false,date=year,backend=biber]{biblatex}
\newcommand{\ket}[1]{\vert{#1}\rangle}
\newcommand{\bra}[1]{\langle{#1}\vert}
\newcommand{\braket}[1]{\langle{#1}\rangle}

\makeatletter
\renewcommand{\@makecaption}[2]{
 \vskip\abovecaptionskip
 \sbox\@tempboxa{#1: #2}{{\bf #1:} #2 \par} 
 \vskip\belowcaptionskip
}
\makeatother
\definecolor{codegreen}{rgb}{0,0.6,0}
\definecolor{gray}{rgb}{0.4,0.4,0.4}
\definecolor{codeblue}{rgb}{0.3,0.5,0.9}
\definecolor{codered}{rgb}{0.78,0.1,0.22}
\definecolor{backcolour}{rgb}{0.95,0.95,0.95}
\lstdefinestyle{python}{
    backgroundcolor=\color{backcolour},   
    commentstyle=\color{gray},
    keywordstyle=\bfseries\color{codeblue},
    numberstyle=\color{orange},
    stringstyle=\color{codegreen},
    basicstyle=\ttfamily\footnotesize,
    breakatwhitespace=false,         
    breaklines=true,                 
    captionpos=b,                    
    keepspaces=true,                 
    numbers=left,                    
    numbersep=5pt,                  
    showspaces=false,                
    showstringspaces=false,
    showtabs=false,                  
    tabsize=2,
    literate=*{0}{{\textcolor{codered}{0}}}{1}%
           {1}{{\textcolor{codered}{1}}}{1}%
           {2}{{\textcolor{codered}{2}}}{1}%
           {3}{{\textcolor{codered}{3}}}{1}%
           {4}{{\textcolor{codered}{4}}}{1}%
           {5}{{\textcolor{codered}{5}}}{1}%
           {6}{{\textcolor{codered}{6}}}{1}%
           {7}{{\textcolor{codered}{7}}}{1}%
           {8}{{\textcolor{codered}{8}}}{1}%
           {9}{{\textcolor{codered}{9}}}{1}
}
\lstdefinestyle{plain}{
    backgroundcolor=\color{backcolour},   
    commentstyle=\color{gray},
    keywordstyle=\color{black},
    numberstyle=\color{orange},
    stringstyle=\color{black},
    basicstyle=\ttfamily\footnotesize,
    breakatwhitespace=false,         
    breaklines=true,                 
    captionpos=b,                    
    keepspaces=true,                 
    numbers=left,                    
    numbersep=5pt,                  
    showspaces=false,                
    showstringspaces=false,
    showtabs=false,                  
    tabsize=2,
    literate=*{0}{{\textcolor{black}{0}}}{1}%
           {1}{{\textcolor{black}{1}}}{1}%
           {2}{{\textcolor{black}{2}}}{1}%
           {3}{{\textcolor{black}{3}}}{1}%
           {4}{{\textcolor{black}{4}}}{1}%
           {5}{{\textcolor{black}{5}}}{1}%
           {6}{{\textcolor{black}{6}}}{1}%
           {7}{{\textcolor{black}{7}}}{1}%
           {8}{{\textcolor{black}{8}}}{1}%
           {9}{{\textcolor{black}{9}}}{1}
}
\input{custom_bib}

\bibliography{bib_programming_qc.bib}
\begin{document}
\title{Lecture Notes: Programming Quantum Computers}
\author[1,2]{Madita Willsch}
\author[1]{Dennis Willsch}
\author[1,2,3]{Kristel Michielsen}
\affil[1]{Institute for Advanced Simulation, J\"ulich Supercomputing Centre\\
Forschungszentrum J\"ulich, 52425 J\"ulich, Germany}
\affil[2]{AIDAS, 52425 J\"ulich, Germany}
\affil[3]{RWTH Aachen University, 52056 Aachen, Germany}
\date{}
\maketitle
\thispagestyle{empty}
\vspace{-4em}
\tableofcontents
\vfill
\noindent\rule{\textwidth}{0.4pt}
\noindent
An earlier version of these lecture notes has been published in \cite{Pavarini2021SimulatingCorrelationsWithComputers}.


\section{Introduction}

Quantum computing is a new emerging computer technology. Current quantum computing\index{quantum computing} devices are at a development stage where they are gradually becoming suitable for small real-world applications. This lecture is devoted to the practical aspects of programming such quantum computing devices.

Over the past twenty years, two major paradigms of quantum computing have emerged. The first is the \emph{gate-based model of quantum computing} (also known as \emph{universal quantum computing}), and the second is \emph{quantum annealing} (also known as \emph{adiabatic quantum computing}). From a mathematical point of view, both models have the same computational power, but in practice they operate in a fundamentally different way.

The first part of this lecture focuses on gate-based quantum computers. We will define the basic unit of computation, the quantum bit (qubit), and how a quantum computer processes information. Subsequently, basic quantum circuits (i.e., the \emph{programs} for gate-based quantum computers) are discussed and simulated. Finally, a more complex algorithm called the \emph{quantum approximate optimization algorithm} (QAOA), which is considered to be an approach to address small optimization problems, is introduced and discussed.

In the second part of this lecture, we give an introduction to quantum annealing and discuss how to program a quantum annealer, i.e., the quantum processing unit (QPU) that performs the quantum annealing process.
The introductory part starts with a discussion of discrete optimization problems and a formulation of the particular set of problems that can be solved on currently available quantum annealers.
Subsequently, we describe the working principles of quantum annealers and the architecture of the currently available quantum annealers by D-Wave Systems Inc. We also discuss physical aspects including some limitations.
Finally, we demonstrate how to program a D-Wave quantum annealer by means of some example programs.

\section{Gate-based quantum computing}

This section provides a hands-on introduction to the programming of gate-based quantum computers. After introducing the basic notions of qubits and gates, several examples of quantum circuits are programmed and discussed. These are either fundamental building blocks in disruptive quantum circuit scenarios, such as the quantum Fourier transform in Shor's factoring algorithm \cite{shor94factoring}, or potentially relevant for near-term applications such as the QAOA \cite{Farhi2014QAOA}. In this section, the term \emph{quantum computer} always refers to the gate-based model of quantum computing.

\subsection{Quantum bits and gates}

\subsubsection{Single qubits}
A gate-based quantum computer is designed to process information in terms of quantum bits\index{quantum bits} (qubits).
The word \emph{qubit} is derived from the basic unit of computation in a digital computer, a \emph{binary digit} or \emph{bit}. While a bit in a digital computer can only ever be either 0 or 1, a qubit is a generalization of a bit in the sense that it can also be in a superposition of 0 and 1.

We describe a qubit $\ket\psi$ in terms of two complex numbers $\psi_0,\psi_1\in\mathbb C$,
\begin{equation}
    \label{eq:qubit}
    \ket{\psi} = \psi_0\ket0 + \psi_1\ket1 = \begin{pmatrix} \psi_0 \\ \psi_1 \end{pmatrix},
\end{equation}
which are normalized such that  $\langle\psi\vert\psi\rangle = \vert\psi_0\vert^2+\vert\psi_1\vert^2 = 1$. In the quantum computer model, the notions of 0 and 1 are represented by the standard vectors
\begin{equation}
    \label{eq:01}
    \ket0 = \begin{pmatrix} 1 \\ 0 \end{pmatrix}, \qquad \ket1 = \begin{pmatrix} 0 \\ 1 \end{pmatrix}.
\end{equation}
For the sake of programming quantum computers, these two notations are equivalent. We call $\ket\psi$ the \emph{state vector} of the qubit.

Informally, a complex superposition like Eq.~(\ref{eq:qubit}) is sometimes described as ``0 and 1 at the same time'', although it is important to realize that the notion of time plays no role here. Equation~(\ref{eq:qubit}) is a well-defined mathematical object that completely describes the state of a single qubit.

\begin{figure}[t!]
\centering
\includegraphics[width=0.4\textwidth]{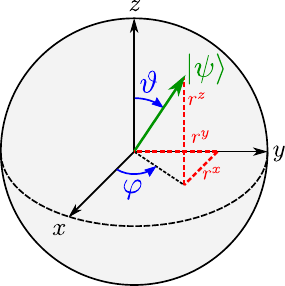}
\caption{Bloch sphere representation of a pure single-qubit state $\ket\psi$.
  The azimuthal angle $\vartheta\in[0,\pi]$ and the polar angle $\varphi\in[0,2\pi)$
  are defined in Eq.~(\ref{eq:singlequbitblochangles}), and the Cartesian coordinates
  $r^x$, $r^y$, and $r^z$ are given by Eq.~(\ref{eq:singlequbitblochvector}).}
\label{fig_michielsen_blochsphere}
\end{figure}

A very useful representation of the general single-qubit state $\ket\psi$ in Eq.~(\ref{eq:qubit}) is called the \emph{Bloch sphere representation} that is shown in Fig.~\ref{fig_michielsen_blochsphere}. It is particularly convenient to visualize the states and operations on a single qubit. We obtain the Bloch sphere\index{Bloch sphere} representation by using the fact that $\langle\psi\vert\psi\rangle = \vert\psi_0\vert^2+\vert\psi_1\vert^2 = 1$, which implies that there exists an angle $\vartheta\in[0,\pi]$ such that $\vert\psi_0\vert=\cos(\vartheta/2)$ and $\vert\psi_1\vert=\sin(\vartheta/2)$. Furthermore, as the global phase of a quantum state is irrelevant, we can choose without loss of generality $\psi_0=\cos(\vartheta/2)$ and $\psi_1=e^{i\varphi}\sin(\vartheta/2)$, where $\varphi\in[0,2\pi)$ represents the relative phase between the complex coefficients. We thus obtain
\begin{equation}
    \label{eq:singlequbitblochangles}
    \ket{\psi} = \cos\frac{\vartheta}2\,\ket0 + e^{i\varphi}\sin\frac{\vartheta}2\,\ket1.
\end{equation}
For all values of $\vartheta$ and $\varphi$, this state can be drawn on the surface of a 3D sphere with radius one as shown in Fig.~\ref{fig_michielsen_blochsphere}.

When a qubit is measured, one always obtains one of the two discrete, digital outcomes ``0'' and ``1''. The complex coefficients of $\ket\psi$ determine the corresponding \emph{probabilities} $p_0=\vert\psi_0\vert^2$ and $p_1=\vert\psi_1\vert^2$ to measure one of the two outcomes. On the Bloch sphere, the probabilities $p_0$ and $p_1$ can be obtained from the projection of $\ket\psi$ onto the $z$ axis.
\\~\\~\\
\textbf{Exercise 1:} Calculate $\vartheta$ and $\varphi$ for the following states, visualize them on the Bloch sphere with radius one, and compute the probabilities to measure the qubit in $\ket0$ and $\ket1$:\\\textbf{(a)} $\ket 0$, \textbf{(c)} $(\ket0+\,\ket1)/\sqrt2$, \textbf{(e)} $(\sqrt3/2)\ket0+((1+i)/2\sqrt2)\ket1$, \\\textbf{(b)} $\ket 1$, \textbf{(d)} $(\ket0+i\ket1)/\sqrt2$, \textbf{(f)} $((1+i)/2)\ket0 + ((1-\sqrt3i)/\sqrt8)\ket1$ (hint: remove the global phase here first).
\\~\\
The Cartesian coordinates $r^x$, $r^y$, and $r^z$ of the single-qubit state $\ket\psi$ in Fig.~\ref{fig_michielsen_blochsphere} can be computed as expectation values of the Pauli matrices\index{Pauli matrices},
\begin{align}
  \sigma^x &= \begin{pmatrix}
    0 & 1 \\ 1 & 0
  \end{pmatrix},&
  \sigma^y &= \begin{pmatrix}
    0 & -i \\ i & 0
  \end{pmatrix},&
  \sigma^z &= \begin{pmatrix}
    1 & 0 \\ 0 & -1
  \end{pmatrix}.
  \label{eq:paulimatrices}
\end{align}
A short calculation yields
\begin{equation}
  \vec{r} =
  \begin{pmatrix}
    r^x \\ r^y \\ r^z
  \end{pmatrix} =
  \begin{pmatrix}
    \braket{\psi|\sigma^x|\psi} \\ \braket{\psi|\sigma^y|\psi} \\ \braket{\psi|\sigma^z|\psi}
  \end{pmatrix} =
  \begin{pmatrix}
    \sin\vartheta\cos\varphi\\
    \sin\vartheta\sin\varphi\\
    \cos\vartheta
  \end{pmatrix}.
  \label{eq:singlequbitblochvector}
\end{equation}

\subsubsection{Quantum gates}
\begin{figure}[t!]
\centering
\includegraphics[width=0.4\textwidth]{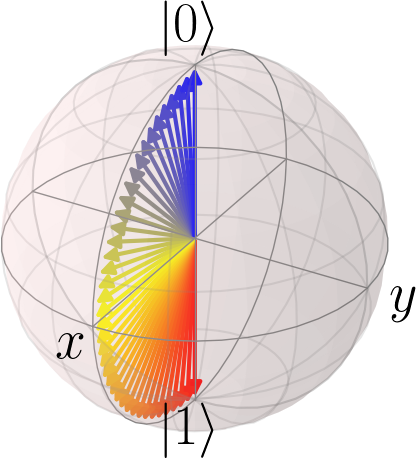}
\caption{Visualization of the single-qubit gate $R^y(\pi)$ (see Eq.~(\ref{eq:singlequbitrotationy})) applied to the state $\ket0$ on the Bloch sphere defined in Fig.~\ref{fig_michielsen_blochsphere}. The gate represents a counter-clockwise rotation around the $y$ axis by an angle of $\pi$. Shown is the time evolution of a qubit during the application of a pulse designed to implement the gate $R^y(\pi)$. The time is encoded in the color of the arrows (from blue at the beginning over yellow to red at the end of the pulse). The data is taken from a simulation of transmon qubits \cite{WillschDennis2020Phd}.}
\label{fig_michielsen_hideal0}
\end{figure}

A \emph{quantum gate} is a unitary operation that can be performed on a qubit. All single-qubit quantum gates\index{quantum gates} can be visualized as rotations of $\ket\psi$ on the Bloch sphere in Fig.~\ref{fig_michielsen_blochsphere}. One defines three elementary qubit rotations as matrix exponentials of the Pauli matrices in Eq.~(\ref{eq:paulimatrices})
\begin{align}
  \label{eq:singlequbitrotationx}
  R^x(\theta) &= e^{-i\theta \sigma^x/2} =
  \begin{pmatrix}
    \cos(\theta/2) & -i\sin(\theta/2) \\
    -i\sin(\theta/2) & \cos(\theta/2)
  \end{pmatrix}
  ,\\
  \label{eq:singlequbitrotationy}
  R^y(\theta) &= e^{-i\theta \sigma^y/2} =
  \begin{pmatrix}
    \cos(\theta/2) & -\sin(\theta/2) \\
    \sin(\theta/2) & \cos(\theta/2)
  \end{pmatrix}
  ,\\
  \label{eq:singlequbitrotationz}
  R^z(\theta) &= e^{-i\theta \sigma^z/2} =
  \begin{pmatrix}
    \exp(-i\theta/2) & 0 \\
    0 & \exp(i\theta/2)
  \end{pmatrix}.
\end{align}
Here, the quantum gate $R^\alpha(\theta)$ for $\alpha=x,y,z$ rotates the qubit $\ket\psi$ by an angle $\theta$ around the axis $\alpha$ according to the \emph{right-hand rule}. This means that if the thumb of the right hand points along the corresponding axis $\alpha$ in Fig.~\ref{fig_michielsen_blochsphere}, the sense of rotation is given by the curl of the remaining fingers, i.e., counter-clockwise when looking at the top of the thumb. An example for the gate $R^y(\pi)$ is shown in Fig.~\ref{fig_michielsen_hideal0}.

Often, at least one of these elementary qubit rotations is implemented in a hardware realization of a gate-based quantum computer. When programming quantum computers, the quantum gates are internally decomposed into products of such elementary rotations. For instance, for the current generation quantum processors of the IBM Q (which are based on superconducting transmon qubits \index{transmon qubits}\cite{koch2007transmon}), the elementary rotations are $R^x(\pi/2)$ and $R^z(\theta)$ \cite{McKay2016VZgate,CROS17}.

A general single-qubit rotation by an angle $\theta$ around a unit axis $\vec n=(n^x,n^y,n^z)$ is given by
\begin{equation}
    \label{eq:singequbitrotationgeneral}
    R^{\vec n}(\theta) = e^{-i\theta \vec n\cdot\vec\sigma/2} = \cos\frac{\theta}2\,I -i \sin\frac{\theta}2\,\vec n\cdot\vec \sigma,
\end{equation}
where $\vec n\cdot\vec\sigma=n^x\sigma^x+n^y\sigma^y+n^z\sigma^z$, and $I$ is the $2\times2$ identity matrix. All single-qubit gates can be written in this form, up to an arbitrary global phase factor of the form $e^{i\alpha}$.

Besides the elementary single-qubit rotations, there are six other important gates that belong to the so-called standard gate set\index{standard gate set}:
\begin{align}
  \label{eq:standardgates1}
  X &= \sigma^x, & Y &= \sigma^y, & Z &= \sigma^z, & \\
  \label{eq:standardgates2}
  H &= \frac{1}{\sqrt2}\begin{pmatrix}
    1 & \phantom{-}1 \\ 1 & -1
  \end{pmatrix},&
  S &= \begin{pmatrix}
    1 & 0 \\ 0 & i
  \end{pmatrix},&
  T &= \begin{pmatrix}
    1 & 0 \\ 0 & e^{i\pi/4}
  \end{pmatrix}.
\end{align}
In particular, the $X$ gate (also known as the NOT gate or \emph{bit flip} gate) and the Hadamard gate $H$ (which maps a state $\ket 0$ to a uniform superposition of $\ket 0$ and $\ket 1$ and back) are used in wide range of applications. A comprehensive list of common quantum gates is given in Appendix~\ref{app:JUQCS}.

As each quantum gate $U$ is unitary (i.e., $U^{-1}=U^\dagger$ where $U^\dagger$ denotes the Hermitian conjugate), the inverse of a quantum gate $U^\dagger$ is also a quantum gate. 
\\~\\
\textbf{Exercise 2:} Find the corresponding axes $\vec n$ and angles $\theta$ (and optionally the global phase factors $e^{i\alpha}$) for all single-qubit gates in Eqs.~(\ref{eq:standardgates1}) and (\ref{eq:standardgates2}), as well as their inverses, and express them in the form of Eq.~(\ref{eq:singequbitrotationgeneral}). Additionally, visualize their operations as rotations on the Bloch sphere, as done in Fig.~\ref{fig_michielsen_hideal0}.

\subsubsection{Multiple qubits}

While a single-qubit state is described by two complex coefficients $\psi_0$ and $\psi_1$ (see Eq.~(\ref{eq:qubit})), a multi-qubit state $\ket\psi$ with $n>1$ qubits is described by $2^n$ complex coefficients $\psi_0, \ldots, \psi_{2^n-1}$,
\begin{equation}
    \label{eq:multiqubit}
    \ket{\psi} = \psi_0\ket{0\cdots00} + \psi_1\ket{0\cdots01} + \cdots + \psi_{2^n-1}\ket{1\cdots11} = \begin{pmatrix} \psi_0 \\ \psi_1 \\ \vdots \\ \psi_{2^n-1} \end{pmatrix},
\end{equation}
The corresponding basis vectors $\ket{q_0q_1\cdots q_{n-1}}$ for $q_i=0,1$ and $i=0,\ldots,n-1$ are constructed from the single-qubit standard basis in Eq.~(\ref{eq:01}) by means of the \emph{tensor product}\index{tensor product} ``$\otimes$'' (also known as \emph{Kronecker product}), $\ket{q_0q_1\cdots q_{n-1}}=\ket{q_0}\otimes\ket{q_1}\otimes\cdots\otimes\ket{q_{n-1}}$. For simplicity, we often do not write the tensor product explicitly.
Consequently, for two qubits, the computational basis reads
\begin{equation}
    \label{eq:twoqubitbasis}
    \ket{00} = \begin{pmatrix} 1 \\ 0 \\ 0 \\ 0\end{pmatrix}, \qquad 
    \ket{01} = \begin{pmatrix} 0 \\ 1 \\ 0 \\ 0\end{pmatrix}, \qquad
    \ket{10} = \begin{pmatrix} 0 \\ 0 \\ 1 \\ 0\end{pmatrix}, \qquad 
    \ket{11} = \begin{pmatrix} 0 \\ 0 \\ 0 \\ 1\end{pmatrix}.
\end{equation}
One may notice that in Eq.~(\ref{eq:multiqubit}), the basis state $\ket{q_0q_1\cdots q_{n-1}}$ corresponding to the coefficient $\psi_j$ for $j=0,\ldots,2^n-1$ contains the binary representation of $j$, i.e., $\mathrm{bin}(j) = q_0q_1\cdots q_{n-1}$, or equivalently, $j=\sum_{i=0}^{n-1} q_i\times2^{n-i-1}$. For this reason, we identify the notations $\ket{j}\equiv\ket{\mathrm{bin}(j)}\equiv\ket{q_0q_1\cdots q_{n-1}}$ so that the state in Eq.~(\ref{eq:multiqubit}) is also written as
\begin{equation}
    \label{eq:multiqubitinteger}
    \ket{\psi} = \sum_{j=0}^{2^n-1} \psi_j\ket j.
\end{equation}
This notation is needed for the example of the quantum Fourier transform discussed below.

Quantum gates on multiple qubits are, like single-qubit gates, unitary operations on the multi-qubit state $\ket\psi$. In practice, most multi-qubit gates are actually single-qubit gates acting on certain qubits in the multi-qubit state. For instance, a single-qubit gate $U$ from Eqs.~(\ref{eq:standardgates1}) and (\ref{eq:standardgates2}) acting on a certain qubit $i$ (denoted by $U_i$) transforms a basis vector $\ket{q_0\cdots q_{n-1}}$ according to
\begin{equation}
    U_i \ket{q_0\cdots q_{n-1}} = \ket{q_0\cdots q_{i-1}}(U\ket{q_i})\ket{q_{i+1}\cdots q_{n-1}}.
\end{equation}
In other words, $U_i$ is given by the tensor product $U_i=I\otimes\cdots\otimes U\otimes\cdots\otimes I$.

Another common set of multi-qubit gates derived from single-qubit gates are \emph{controlled} quantum gates\index{controlled quantum gate}. For a single-qubit gate $U$, the controlled-$U$ gate (denoted by C$U$) acts on two qubits $i_1$ and $i_2$ in a multi-qubit state. Its action on a basis vector $\ket{q_0\cdots q_{n-1}}$ is defined by
\begin{equation}
    \label{eq:controlledU}
    \mathrm{C}U_{i_1i_2} \ket{q_0\cdots q_{n-1}} = \begin{cases}
        \hfil\ket{q_0\cdots q_{n-1}} & \text{(if $q_{i_1}=0$)} \\
        \ket{q_0\cdots q_{i_2-1}}(U\ket{q_{i_2}})\ket{q_{i_2+1}\cdots q_{n-1}} & \text{(if $q_{i_1}=1$)}
    \end{cases}.
\end{equation}
In other words, the action is controlled by qubit $q_{i_1}$, i.e., the single-qubit gate $U$ only acts on the target qubit $q_{i_2}$ if the control qubit $q_{i_1}$ is in state 1. 
On the two-qubit space spanned by the basis states in Eq.~(\ref{eq:twoqubitbasis}), the matrix representation of the controlled-$U$ gate is given by 
\begin{equation}
    \label{eq:controlledUtwoqubits}
    \mathrm{C}U = \begin{pmatrix}
      I & \mathbf0 \\
      \mathbf0 & U
    \end{pmatrix},
\end{equation}
where $\mathbf0$ denotes a $2\times2$ matrix with all elements equal to zero. 

It is important to realize that for controlled quantum gates constructed in this way, the global phase of the single-qubit gate $U$ becomes a relative phase. In particular, this means that, even though the single-qubit gates $S$ and $R^z(\pi/2)$ are equivalent, the controlled gates C$S$ and C$R^z(\pi/2)$ are different two-qubit gates.

Two very important two-qubit gates constructed like this are the controlled-NOT (CNOT or C$X$) and the controlled-phase (C$Z$) gates. Their matrix representations with respect to the two-qubit basis in Eq.~(\ref{eq:twoqubitbasis}) are given by
\begin{align}
    \mathrm{CNOT} &= \begin{pmatrix}
      1 & 0 & 0 & 0 \\
      0 & 1 & 0 & 0 \\
      0 & 0 & 0 & 1 \\
      0 & 0 & 1 & 0 \\
    \end{pmatrix},
    &
    \mathrm{C}Z &= \begin{pmatrix}
      1 & 0 & 0 & 0 \\
      0 & 1 & 0 & 0 \\
      0 & 0 & 1 & 0 \\
      0 & 0 & 0 & -1 \\
    \end{pmatrix}.
\end{align}
The CNOT gate and the C$Z$ gate can be converted into one another using the identity
$\mathrm{CNOT} = (I\otimes H)\,\mathrm{C}Z\,(I\otimes H)$, which can be verified by computing the product of the corresponding matrix representations. On a space with more than two qubits, the same identity reads $\mathrm{CNOT}_{i_1i_2} = H_{i_2}\,\mathrm{C}Z_{i_1i_2}\,H_{i_2}$. More of such \emph{circuit identities}\index{quantum circuit identities} that are useful when programming quantum computers can be found in the following exercise and in \cite{NIEL10}.
\\~\\
\textbf{Exercise 3:} Verify the following circuit identities, e.g. by computing their matrix representations on a suitable space and then confirming that they are equivalent (up to a global phase):
\begin{tabular}{lll}
    \textbf{(a)} $X=HZH$ & \textbf{(b)} $H=SR^x(\pi/2)S$ & \textbf{(c)} $I=XX=YY=ZZ=HH$\\
    \textbf{(d)} $Y=H\,YH$ & \textbf{(e)} $XR^y(\theta)X=R^y(-\theta)$ & \textbf{(f)} $\mathrm{CNOT}_{i_2i_1} = H_{i_1}H_{i_2}\,\mathrm{CNOT}_{i_1i_2}\,H_{i_1}H_{i_2}$ \\
    \textbf{(g)} $Z=HXH$ & \textbf{(h)} $HTH=R^x(\pi/4)$ & \textbf{(i)} $\mathrm{CNOT}_{i_2i_1} = H_{i_1}H_{i_2}\,\mathrm{CNOT}_{i_1i_2}\,H_{i_1}H_{i_2}$ \\
    \textbf{(j)} $\mathrm{C}Z_{i_1i_2} = \mathrm{C}Z_{i_2i_1}$ & \textbf{(k)} $\mathrm{C}(e^{i\alpha}I)=R^z(\alpha)\otimes I$ &  \textbf{(l)} $R^z_{i_1}(\theta)\,\mathrm{CNOT}_{i_1i_2} = \mathrm{CNOT}_{i_1i_2}R^z_{i_1}(\theta)$ \\
    \textbf{(m)} $\mathrm{C}S_{i_1i_2} = \mathrm{C}S_{i_2i_1}$ & \textbf{(n)} $\mathrm{CNOT}_{i_1i_2} = H_{i_2}\,\mathrm{C}Z_{i_1i_2}\,H_{i_2}$ & \textbf{(o)} $R^x_{i_2}(\theta)\,\mathrm{CNOT}_{i_1i_2} = \mathrm{CNOT}_{i_1i_2}R^x_{i_2}(\theta)$ \\
    \textbf{(p)} $S=TT$
\end{tabular}

\subsection{Programming and simulating quantum circuits}

\begin{figure}[t!]
\centering
\begin{tabular}{ccccccc}
    \raisebox{2em}{(a)}
    &
    \begin{tabular}{l}
        $\Qcircuit @C=1.0em @R=0.5em @!R {
                 & \gate{H} & \gate{T} & \gate{H} & \ctrl{1} & \qw \\
                 & \qw & \qw & \qw & \targ & \qw
         }$
    \end{tabular}
    &
    \raisebox{2em}{(b)}
    &
    \begin{tabular}{l}
        $\Qcircuit @C=1.0em @R=0.5em @!R {
                 & \ctrl{1} & \qw \\
                 & \gate{S} & \qw
         }$
    \end{tabular}
    &
    $=$
    &
    \begin{tabular}{l}
        $\Qcircuit @C=1.0em @R=0.5em @!R {
                     & \qw & \ctrl{1} & \qw & \ctrl{1} & \gate{T} & \qw\\
                     & \gate{T} & \targ & \gate{T^\dag} & \targ & \qw & \qw
             }$
    \end{tabular}
\end{tabular}
\caption{Example quantum circuit diagrams. Note that the diagrams are read from left to right. (a) Quantum circuit that generates the state $\cos(\pi/8)\ket{00}-i\sin(\pi/8)\ket{11}$ (which can be computed using the circuit identity from Exercise 3 (h)) when starting with the initial state $\ket{00}$ on the left. (b) Schematic way of writing a circuit identity for the C$S$ gate.}
\label{fig_michielsen_CS}
\end{figure}
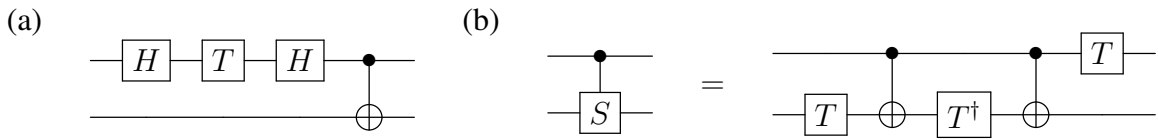

A program for a gate-based quantum computer is called a \emph{quantum circuit}\index{quantum circuit}. A quantum circuit is a sequence of multiple quantum gates. It is often expressed in a diagrammatic language that uses horizontal lines to represent the qubits and boxes to represent the quantum gates. The order of execution in the quantum gate sequence is from left to right. Controlled quantum gates such as C$U_{i_1i_2}$ in Eq.~(\ref{eq:controlledU}) are visualized with a filled dot on the control qubit and the single-qubit gate $U$ on the target qubit. The CNOT gate in particular is visualized with an encircled plus symbol on the target qubit. Two examples using this schematic language are shown in Fig.~\ref{fig_michielsen_CS}.

As described in the previous section, each gate in a quantum circuit represents a unitary matrix. By multiplying all quantum gate matrices in a quantum circuit, we could, in principle, obtain a large unitary matrix that is equivalent to the quantum circuit. Note that, as the order of execution in a quantum circuit diagram is from left to right, the corresponding quantum gate matrices must be multiplied in reverse order. Take for instance the right circuit in Fig.~\ref{fig_michielsen_CS}(a), which has the corresponding matrix product $\mathrm{CNOT}\,(H\otimes I)\,(T\otimes I)\,(H\otimes I)$.

To simulate a quantum circuit, one could use the thus obtained matrix and apply it to a certain initial state, which is typically chosen as $\ket{0\cdots0} = (1,0,\ldots,0)^T$. This method works for small quantum circuits. However, it quickly becomes prohibitive because the size of the quantum circuit matrix for $N$ qubits is $2^N\times 2^N$.

Larger quantum circuits with, say, $N\ge30$ qubits can be simulated on supercomputers such as the GPU-cluster JUWELS Booster using JUQCS--G \cite{Willsch2021JUQCSGQAOA}, which is a GPU-accelerated version of the J\"ulich Universal Quantum Computer Simulator (JUQCS) \cite{RAED07x,RAED19a} that was also used for benchmarking purposes in Google's quantum supremacy experiment \cite{Google2019QuantumSupremacy}. For reference, the standard gate set implemented by JUQCS is given in Appendix~\ref{app:JUQCS}. 

A \texttt{qiskit} \cite{Qiskit} interface to JUQCS including the conversion from the \texttt{qiskit} gate set to the JUQCS gate set is available through the J\"ulich UNified Infrastructure for Quantum computing (JUNIQ) service at \url{https://jugit.fz-juelich.de/qip/juniq-platform}.

An example program to simulate the circuit in Fig.~\ref{fig_michielsen_CS}(a) using this interface is shown in Listing~\ref{juniqexample1}. It simulates the quantum circuit by propagating the state vector, sampling from the resulting probability, and returning the counts for all sampled events (in this case ``00'' and ``11''). Furthermore, instead of sampling, JUQCS also provides an option to return the full state vector up to a certain number of qubits. An example program for this mode of operation is shown in Listing~\ref{juniqexample2}.

\begin{lstlisting}[float, language=Python, caption={Example program to simulate the quantum circuit shown in Fig.~\ref{fig_michielsen_CS}(a). As the result of this circuit is $\cos(\pi/8)\ket{00}-i\sin(\pi/8)\ket{11}$, the printed counts should correspond to the probabilities $p_{00}=\cos(\pi/8)^2\approx0.85$ and  $p_{11}=\sin(\pi/8)^2\approx0.15$.  Note also the usage of \texttt{circuit.reverse\_bits()}, because \texttt{qiskit} uses the ordering $\ket{q_{n-1}\cdots q_{0}}$ while all standard text books as well as these lecture notes use $\ket{q_0\cdots q_{n-1}}$.}, label=juniqexample1]
# load the qiskit module
import qiskit

# create a quantum circuit with 2 qubits
circuit = qiskit.QuantumCircuit(2)
# add gates to the quantum circuit acting on qubits 0 and 1
circuit.h(0)
circuit.t(0)
circuit.h(0)
circuit.cx(0,1)
# measure all qubits
circuit.measure_all()

# load JUQCS
from juqcs import Juqcs
# use JUQCS as backend
backend = Juqcs.get_backend('qasm_simulator')
# allocate 10 minutes on the supercomputer with memory for a 2 qubit circuit
backend.allocate(minutes=10, max_qubits=2)

# run the circuit on the supercomputer
job = qiskit.execute(circuit.reverse_bits(), backend=backend, shots=1000)
# store the result
result = job.result()

# print the part you are interested in (here the counts)
print(result.get_counts())

# deallocate the reserved resources on the supercomputer
backend.deallocate()
\end{lstlisting}

\begin{lstlisting}[float, language=Python, caption={Example program to simulate the quantum circuit shown in Fig.~\ref{fig_michielsen_CS}(a) using the state vector simulator. The result of this program is a numerical representation of the state $\cos(\pi/8)\ket{00}-i\sin(\pi/8)\ket{11}$, up to a global phase. Note also the usage of \texttt{circuit.reverse\_bits()}, because \texttt{qiskit} uses the ordering $\ket{q_{n-1}\cdots q_{0}}$ while all standard text books as well as these lecture notes use $\ket{q_0\cdots q_{n-1}}$.}, label=juniqexample2]
import qiskit

circuit = qiskit.QuantumCircuit(2)
circuit.h(0)
circuit.t(0)
circuit.h(0)
circuit.cx(0,1)

from juqcs import Juqcs
backend = Juqcs.get_backend('statevector_simulator')
backend.allocate(minutes=10, max_qubits=2)

job = qiskit.execute(circuit.reverse_bits(), backend=backend)
result = job.result()

print(result.get_statevector())

backend.deallocate()
\end{lstlisting}

Instead of the JUQCS backend, one can also use a simulator from \texttt{qiskit} via the function \texttt{qiskit.Aer.get\_backend}. This one only works for circuits with a small number of qubits, but it does not need the calls to \texttt{backend.allocate} and  \texttt{backend.deallocate} as it does not run on a supercomputer. Furthermore, backends for real quantum devices can be accessed in a similar manner through the IBM Q Experience \cite{ibmquantumexperience2016}.

\subsection{Example: Quantum adder}
\index{quantum adder}

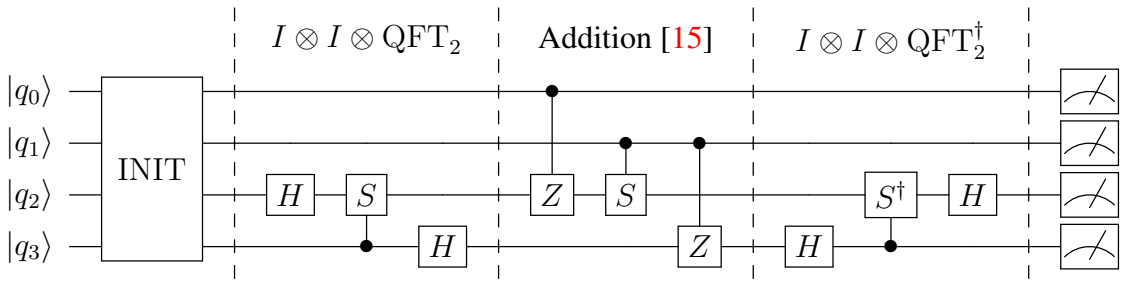
\begin{figure}[t!]
\centering
\begin{equation*}
    \Qcircuit @C=1.0em @R=0.2em @!R {
                & \barrier[0em]{4} & & & \mbox{$I\otimes I\otimes\mathrm{QFT}_{2}$} & \barrier[0em]{4} & & & \mbox{Addition \cite{DRAP00}} & \barrier[0em]{4} & & & \mbox{$I\otimes I\otimes\mathrm{QFT}_{2}^\dagger$} & \barrier[0em]{4} & \\
                \lstick{ \ket{{q}_{0}}  } & \multigate{3}{\mathrm{INIT}} & \qw & \qw & \qw & \qw & \qw & \ctrl{2} & \qw & \qw & \qw & \qw & \qw & \qw & \qw & \meter\\
                \lstick{ \ket{{q}_{1}}  } & \ghost{\mathrm{INIT}} & \qw & \qw & \qw & \qw & \qw & \qw & \ctrl{1} & \ctrl{2} & \qw & \qw & \qw & \qw & \qw & \meter\\
                \lstick{ \ket{{q}_{2}}  } & \ghost{\mathrm{INIT}} & \qw & \gate{H} & \gate{S} & \qw & \qw & \gate{Z} & \gate{S} & \qw & \qw & \qw & \gate{S^\dag} & \gate{H} & \qw & \meter \\
                \lstick{ \ket{{q}_{3}}  } & \ghost{\mathrm{INIT}} & \qw & \qw & \ctrl{-1} & \gate{H} & \qw & \qw & \qw & \gate{Z} & \qw & \gate{H} & \ctrl{-1} & \qw & \qw & \meter\\
         }
\end{equation*}
\caption{Circuit for a quantum adder \cite{DRAP00} that adds two two-qubit registers modulo four, according to the rule Eq.~(\ref{eq:adder}). The circuit consists of four parts: Initialization of the qubit registers, QFT on the last two registers, the phase addition transform from \cite{DRAP00}, and another QFT on the last two registers. Note that the swaps that are part of typical QFT circuits (see \cite{NIEL10}) are left out. Finally, a measurement of the qubits, which produces a ``0'' or a ``1'' for each qubit, is indicated at the end. To rewrite the gates, one can use quantum circuit identities from Exercise 3 or \cite{NIEL10}.}
\label{fig_michielsen_adder_bare}
\end{figure}

As a first ``real-world'' example, we consider a four-qubit quantum computer program that adds two two-qubit registers according to the rule
\begin{equation}
    \label{eq:adder}
    \ket{q_0q_1}\ket{q_2q_3} \mapsto \ket{q_0q_1}\ket{q_0q_1 + q_2q_3},
\end{equation}
where the expression $q_0q_1 + q_2q_3$ is to be understood as integer addition modulo four (i.e., the result is wrapped back into the range $0,1,2,3$) of the two-bit integers with binary representations $q_0q_1$ and $q_2q_3$, respectively. After the application of the quantum circuit, the second register contains the sum in binary notation. Some examples of the operation of the quantum adder are:
\begin{align}
    \label{eq:adderex1}
    &\ket{2}\ket{1} &&\mapsto &&\ket{2}\ket{3}, &\\
    \label{eq:adderex2}
    &\ket{2}\frac{\ket0+\ket1}{\sqrt2} &&\mapsto &&\ket{2}\frac{\ket2+\ket3}{\sqrt2},\\
    \label{eq:adderex3}
    &\ket{2}\frac{\ket0+\ket1+\ket2}{\sqrt3} &&\mapsto &&\ket{2}\frac{\ket2+\ket3+\ket0}{\sqrt3}.
\end{align}
The interesting thing is that it can also add superpositions in parallel, as done in Eqs.~(\ref{eq:adderex2}) and (\ref{eq:adderex3}) (note that this works because quantum circuits are linear maps, so a circuit $U$ applied to a state $\ket{\psi_1}+\ket{\psi_2}$ produces the state $U\ket{\psi_1}+U\ket{\psi_2}$).
\\~\\
\textbf{Exercise 3:} Use the rule Eq.~(\ref{eq:adder}) for the quantum adder to work out the result when the initial state is given by $(\ket0+\ket3)/\sqrt{2}\otimes(\ket1+\ket2+\ket 3)/\sqrt{3}$.
\\~\\
The quantum circuit for the adder consists of four stages that are shown in Fig.~\ref{fig_michielsen_adder_bare}. The purpose of the INIT block at the beginning is to initialize the registers in a certain initial state, such as the states on the left-hand side in Eqs.~(\ref{eq:adderex1})--(\ref{eq:adderex3}).

The second and the fourth stage contain a quantum Fourier transform\index{quantum Fourier transform} (QFT) on the second register. The QFT is an operation that, loosely speaking, moves information from the registers to the phases of exponential prefactors and vice versa. For an initial $N$-qubit basis state $\ket j$ ($j=0,\ldots,2^N-1$), the QFT is defined as the unitary transformation
\begin{align}
  \label{eq:multiqubitqft}
  \ket j &\stackrel{\mathrm{QFT}}{\mapsto} \frac{1}{2^{N/2}} \sum_{k=0}^{2^N-1} e^{2\pi i j k /2^N} \ket k,
\end{align}
where we identified $j$ and its binary representation $q_0\cdots q_{N-1}$ as
done in Eq.~(\ref{eq:multiqubitinteger}). There is a generic quantum circuit \cite{NIEL10} to implement this transformation using only $H$ gates and conditional $z$ rotations in a
number of steps \emph{polynomial} in $N$, as opposed to the \emph{Fast Fourier
Transform} that requires $\mathcal O(N2^N)$ steps. It is an important component of many quantum algorithms for which a theoretical exponential speedup is known. One such algorithm is Shor's factorization algorithm \cite{shor94factoring} in
which the QFT is basically used to find the period of a suitable function  (note
that finding periods is a generic feature of any Fourier transform).

The third part of the quantum adder circuit in Fig.~\ref{fig_michielsen_adder_bare} is the addition transform from \cite{DRAP00}. It uses conditional phase shifts from the first to the second register so that an addition is effectively performed in the exponent of the phase factors. For instance, if the first register represents an integer $l$ (e.g.~$\ket l\equiv\ket{q_0q_1}$ above), and the second register is given by the QFT of an integer $j$ (i.e.~$\mathrm{QFT} \ket{j}\propto\sum_k e^{2\pi ijk/2^N}\ket k$), the addition transform performs the operation
\begin{align}
  \label{eq:multiqubitadder}
  \sum_{k=0}^{2^N-1} e^{2\pi i j k /2^N} \ket l\ket k \mapsto \sum_{k=0}^{2^N-1} e^{2\pi i (j+l) k /2^N} \ket l\ket k.
\end{align}
After this step, the inverse QFT (given by $\mathrm{QFT}^\dagger$ in Fig.~\ref{fig_michielsen_adder_bare}) can be used to move the result $(j+l)$ of the addition from the exponent back into the second register. Note that the addition is automatically implemented modulo $2^N$, because the complex exponential function is periodic with period $2\pi i$. 

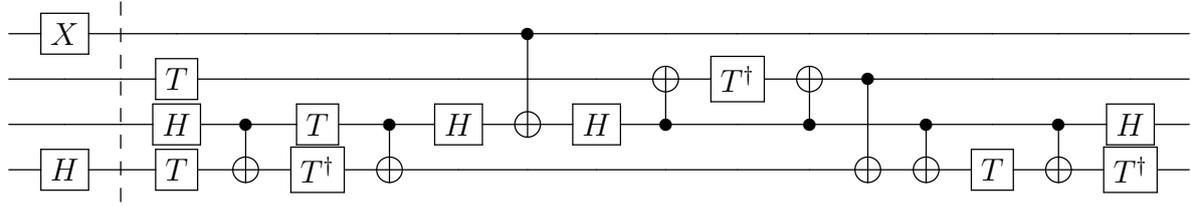
\begin{figure}[p!]
\centering
\begin{equation*}
    \Qcircuit @C=1.0em @R=0.0em @!R {
                & \gate{X} \barrier[0em]{3} & \qw & \qw & \qw & \qw & \qw & \qw & \ctrl{2} & \qw & \qw & \qw & \qw & \qw & \qw & \qw & \qw & \qw & \qw\\
                & \qw & \qw & \gate{T} & \qw & \qw & \qw & \qw & \qw & \qw & \targ & \gate{T^\dag} & \targ & \ctrl{2} & \qw & \qw & \qw & \qw & \qw\\
                & \qw & \qw & \gate{H} & \ctrl{1} & \gate{T} & \ctrl{1} & \gate{H} & \targ & \gate{H} & \ctrl{-1} & \qw & \ctrl{-1} & \qw & \ctrl{1} & \qw & \ctrl{1} & \gate{H} & \qw\\
                & \gate{H} & \qw & \gate{T} & \targ & \gate{T^\dag} & \targ & \qw & \qw & \qw & \qw & \qw & \qw & \targ & \targ & \gate{T} & \targ & \gate{T^\dag} & \qw\\
         }
\end{equation*}
\caption{Example circuit for the quantum adder in Fig.~\ref{fig_michielsen_adder_bare} after transpiling some of the quantum gates using appropriate circuit identities. Note that in this example, an $X$ and an $H$ gate have been used to explicitly create the initial state for the example in Eq.~(\ref{eq:adderex2}).}
\label{fig_michielsen_adder}
\end{figure}

As an example, we consider an implementation of the quantum adder using the single-qubit gate set defined in Eqs.~(\ref{eq:singequbitrotationgeneral})--(\ref{eq:standardgates2}) and the CNOT gate as the only two-qubit gate. This requires rewriting (also known as \emph{transpiling}) the gates using the circuit identites from Exercise 3 and Fig.~\ref{fig_michielsen_CS}(b). The result is shown in Fig.~\ref{fig_michielsen_adder} and Listing~\ref{juniqexampleadder}. Note that the initial state in this example is chosen to be $\ket2(\ket0+\ket1)/\sqrt2$, i.e., the example from Eq.~(\ref{eq:adderex2}) above. As the output contains a superposition of the results, each result occurs as separate events in the simulation.

\begin{lstlisting}[float, language=Python, caption={Example program for the quantum adder circuit shown in Fig.~\ref{fig_michielsen_adder} for the case given by Eq.~(\ref{eq:adderex2}) with initial state $\ket2(\ket0+\ket1)/\sqrt2$. The circuit is simulated with JUQCS as shown in Listing~\ref{juniqexample1}. The output should be the events ``1010'' (corresponding to state $\ket2\ket2$) and ``1011'' (corresponding to state $\ket2\ket3$) with roughly 50\% frequency each.}, label=juniqexampleadder]
import qiskit

circuit = qiskit.QuantumCircuit(4)
circuit.x(0)
circuit.h(3)
circuit.barrier()
circuit.t(1)
circuit.h(2)
circuit.t(3)
circuit.cx(2,3)
circuit.tdg(3)
circuit.t(2)
circuit.cx(2,3)
circuit.h(2)
circuit.cx(0,2)
circuit.h(2)
circuit.cx(2,1)
circuit.tdg(1)
circuit.cx(2,1)
circuit.cx(1,3)
circuit.cx(2,3)
circuit.t(3)
circuit.cx(2,3)
circuit.h(2)
circuit.tdg(3)
circuit.measure_all()

from juqcs import Juqcs
backend = Juqcs.get_backend('qasm_simulator')
backend.allocate(minutes=10, max_qubits=circuit.num_qubits)

job = qiskit.execute(circuit.reverse_bits(), backend=backend, shots=1000)
result = job.result()

print(result.get_counts())

backend.deallocate()
\end{lstlisting}

\subsection{Example: Quantum approximate optimization algorithm}
\index{quantum approximate optimization algorithm}

\begin{figure}[t!]
\centering
\begin{equation*}
    \Qcircuit @C=1.2em @R=0.7em @!R {
                & & & & \mbox{Repeat for $k=1,\ldots,p$ QAOA steps \hphantom{MM}} & & \\
                & & & & \mbox{Weighting \hphantom{MMMMMMMM}} & \mbox{Mixing} & & \\
                \lstick{ \ket{{q}_{0}}  } & \gate{H} & \qw & \gate{R^z(2\gamma_kh_0)} & \multigate{3}{\prod\limits_{i,j} e^{-i\gamma_k J_{ij}\sigma_i^z\sigma_j^z}} & \gate{R^x(2\beta_k)} & \qw & \meter \\
                \lstick{ \ket{{q}_{1}}  } & \gate{H} & \qw & \gate{R^z(2\gamma_kh_1)} & \ghost{\prod\limits_{i,j} e^{-i\gamma_k J_{ij}\sigma_i^z\sigma_j^z}} & \gate{R^x(2\beta_k)} & \qw & \meter\\
                \lstick{ \vdots\:\:\,  } & \vdots & & \vdots & & \vdots & & \vdots \\
                \lstick{ \ket{{q}_{N-1}}  } & \gate{H} & \qw & \gate{R^z(2\gamma_kh_{N-1})} & \ghost{\prod\limits_{i,j} e^{-i\gamma_k J_{ij}\sigma_i^z\sigma_j^z}} & \gate{R^x(2\beta_k)} & \qw & \meter \gategroup{3}{4}{6}{5}{1em}{^\}} \gategroup{3}{6}{6}{6}{1em}{^\}}\gategroup{2}{4}{6}{6}{1.5em}{--}  \\
         }
\end{equation*}
\caption{General QAOA circuit \cite{Farhi2014QAOA}. Initially, the qubits are brought into a uniform superposition over all states using $H$ gates. Then, $k=1,\ldots,p$ QAOA steps with variational parameters $\beta_1,\ldots,\beta_p$ and $\gamma_1,\ldots,\gamma_p$ are applied. Each QAOA step $k$ consists of a ``weighting step'' using $z$ rotations scaled with $\gamma_k$ and the parameters $\{h_i\}$ and $\{J_{ij}\}$ that define the optimization problem (see Eq.~(\ref{eq:qaoaenergy})), followed by a ``mixing step'' using $x$ rotations with angle $2\beta_k$. Finally, the qubits are measured. The result can then be used to update the variational parameters until the energy is sufficiently low.}
\label{fig_michielsen_qaoa}
\end{figure}
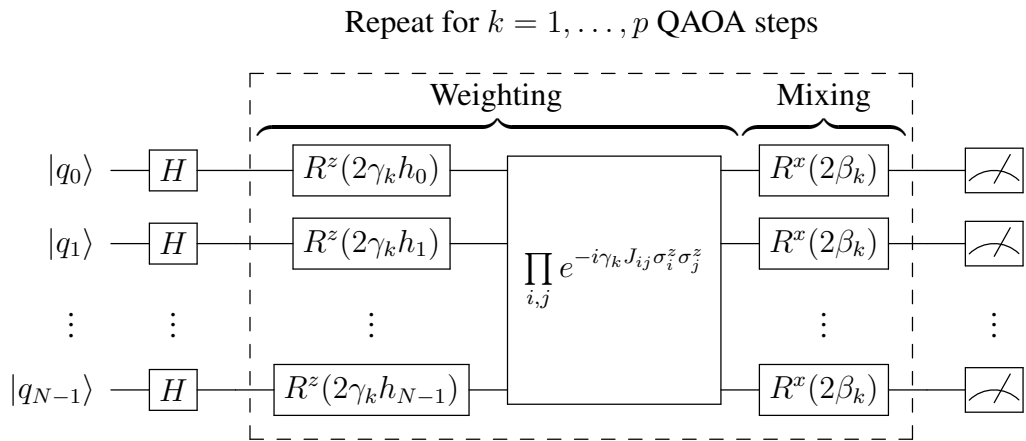

In this section, we consider an approach to address optimization problems on gate-based quantum computers.
Gate-based quantum computers are not made to solve optimization problems by design, which is a key difference to the adiabatic quantum computers covered in the next section. However, there is a standard (and by now rather common) way of addressing optimization problems in QUBO/Ising form on gate-based quantum computers, namely the quantum approximate optimization algorithm (QAOA) \cite{Farhi2014QAOA}.

The optimization problems considered here are discrete Ising problems. The goal of such problems is to find the minimum of an energy function (or cost function)
\begin{equation}
  \label{eq:qaoaenergy}
  E(s_0,\ldots,s_{N-1}) = \sum\limits_{i=0}^{N-1}h_i s_i + \sum\limits_{i<j}J_{ij}s_is_j,
\end{equation}
where $s_0,\ldots,s_{N-1}=\pm1$ are the discrete, two-valued variables, $N$ is the number of variables (which is equal to the number of required qubits), and $\{h_i\}$ and $\{J_{ij}\}$ are real numbers that define the optimization problem instance (for more information see the following section).

The QAOA is a quantum algorithm to find minima (or low-energy states) of Eq.~(\ref{eq:qaoaenergy}). It is a variational quantum algorithm, which means that it has a number of variational parameters $\beta_1,\ldots,\beta_p$ and $\gamma_1,\ldots,\gamma_p$ that are varied during the iterations of the algorithm. The order of the QAOA, denoted by $p$, determines the number of variational parameters. It is worth mentioning that for large $p$, the QAOA can be related to approximate quantum annealing, which also provides a method to find a useful initialization of the variational parameters (see \cite{Willsch2021JUQCSGQAOA}).

\begin{figure}[t!]
\centering
\begin{tabular}{cccc}
    \begin{tabular}{l}
        $\Qcircuit @C=1.0em @R=0.5em @!R {
                 \lstick{ \ket{{q}_{i}}  } & \multigate{1}{e^{-i\gamma_k J_{ij}\sigma_i^z\sigma_j^z}} & \qw \\
                 \lstick{ \ket{{q}_{j}}  } & \ghost{e^{-i\gamma_k J_{ij}\sigma_i^z\sigma_j^z}} & \qw
         }$
    \end{tabular}
    &
    $=$
    &
    \hphantom{M}
    &
    \begin{tabular}{l}
        $\Qcircuit @C=1.0em @R=0.5em {
                     \lstick{ \ket{{q}_{i}}  } & \ctrl{1} & \qw & \ctrl{1} & \qw\\
                     \lstick{ \ket{{q}_{j}}  } & \targ & \gate{R^z(2\gamma_k J_{ij})} & \targ & \qw
             }$
    \end{tabular}
\end{tabular}
\caption{Circuit identity to implement the second part of the weighting step in Fig.~\ref{fig_michielsen_qaoa}.}
\label{fig_michielsen_zz_identity}
\end{figure}
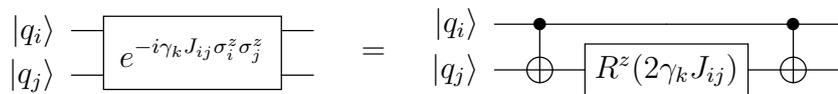

In each iteration of the QAOA, one executes the quantum circuit in Fig.~\ref{fig_michielsen_qaoa} for a given set of variational parameters $\beta_1,\ldots,\beta_p$ and $\gamma_1,\ldots,\gamma_p$ (note that the quantum circuit depends on the fixed problem instance given by $\{h_i\}$ and $\{J_{ij}\}$). To execute the gate $\prod_{i,j} e^{-i\gamma_k J_{ij}\sigma_i^z\sigma_j^z}$ in the weighting step of the QAOA circuit in Fig.~\ref{fig_michielsen_qaoa} using the standard gate set defined above, we need another circuit identity. This identity is shown in Fig.~\ref{fig_michielsen_zz_identity}.
\\~\\
\textbf{Exercise 4:} Prove the circuit identity in Fig.~\ref{fig_michielsen_zz_identity}, e.g.~by computing the matrix representations of both sides and verifying that they are equivalent.
\\~\\
Each measurement at the end of the circuit produces a string of discrete variables $s_0,\ldots,s_{N-1}=\pm1$, where the qubit measurement $q_i=0$ corresponds to $s_i=+1$ and $q_i=1$ corresponds to $s_i=-1$. Note that this convention, $q_i=(1-s_i)/2$, is standard for gate-based quantum computers \cite{NIEL10}; for quantum annealers, one often uses $q_i=(1+s_i)/2$ instead (see below). 

From several executions of the circuit, an expectation value for the energy in Eq.~(\ref{eq:qaoaenergy}) can be computed. This result can be used to update the variational parameters and perform the next iteration. This process is continued as long as necessary until it converges. 

Note that, when the QAOA is simulated using a state vector simulator (see Listing~\ref{juniqexample2}), the expectation value can also be computed directly from the state vector $\ket\psi$. Formally, this can be done by replacing the problem variables $s_i$ in Eq.~(\ref{eq:qaoaenergy}) by the Pauli matrices $\sigma_i^z$ (which yields the \emph{Ising Hamiltonian} of the problem) and computing the expectation value
\begin{equation}
  \label{eq:qaoaenergyexpectation}
  \langle E\rangle = \bra\psi E(\sigma_0^z,\ldots,\sigma_{N-1}^z)\ket\psi.
\end{equation}

As an example, we consider a three-qubit problem characterized by the energy function
\begin{equation}
  \label{eq:qaoaenergyexample}
  E(s_0,s_1,s_2) = - s_0 + \frac 1 2 s_1 - \frac 1 2 s_2 + \frac 1 2 s_0s_1 + \frac 1 2 s_1s_2.
\end{equation}
In this case, the problem parameters for Eq.~(\ref{eq:qaoaenergy}) are given by $(h_0,h_1,h_2)=(-1,1/2,-1/2)$ and $(J_{01},J_{02},J_{12})=(1/2,0,1/2)$. The minimum is given by $(s_0,s_1,s_2)=(+1,-1,+1)$ and corresponds to the state vector $\ket{010}$. The energy at the minimum is $E(+1,-1,+1)=-3$. 

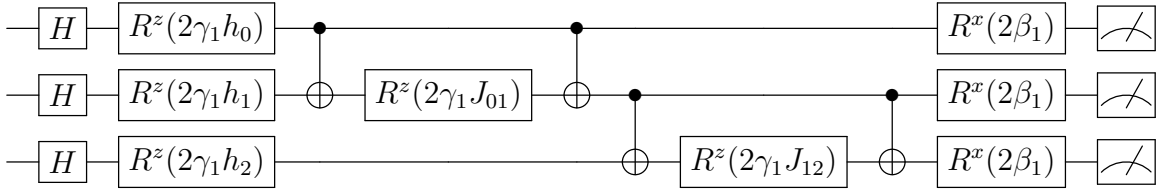
\begin{figure}[t!]
\centering
\begin{equation*}
    \Qcircuit @C=1.0em @R=0.5em {
                &\gate{H} & \gate{R^z(2\gamma_1h_0)} & \ctrl{1} & \qw & \ctrl{1} & \qw & \qw & \qw & \gate{R^x(2\beta_1)} & \meter \\
                &\gate{H} & \gate{R^z(2\gamma_1h_1)} & \targ & \gate{R^z(2\gamma_1 J_{01})} & \targ & \ctrl{1} & \qw & \ctrl{1} & \gate{R^x(2\beta_1)} & \meter\\
                &\gate{H} & \gate{R^z(2\gamma_1h_2)} & \qw & \qw & \qw & \targ & \gate{R^z(2\gamma_1 J_{12})} & \targ & \gate{R^x(2\beta_1)} & \meter \\
         }
\end{equation*}
\caption{Example QAOA circuit for $p=1$ to find the minimum of the energy function given by Eq.~(\ref{eq:qaoaenergyexample}). The circuit has three qubits (one for each problem variable $s_i$). The qubit values $q_i$ after the measurement at the end are related to the problem variables via $q_i=(1-s_i)/2$ (gate-based convention). Note that only two of the blocks in Fig.~\ref{fig_michielsen_zz_identity} are necessary for the weighting step, because the third coupling parameter $J_{02}=0$.}
\label{fig_michielsen_qaoa_example}
\end{figure}

\begin{figure}[t!]
 \centering
 \begin{tabular}{cccc}
    \raisebox{12em}{(a)}
    &
    \includegraphics[width=0.4\textwidth]{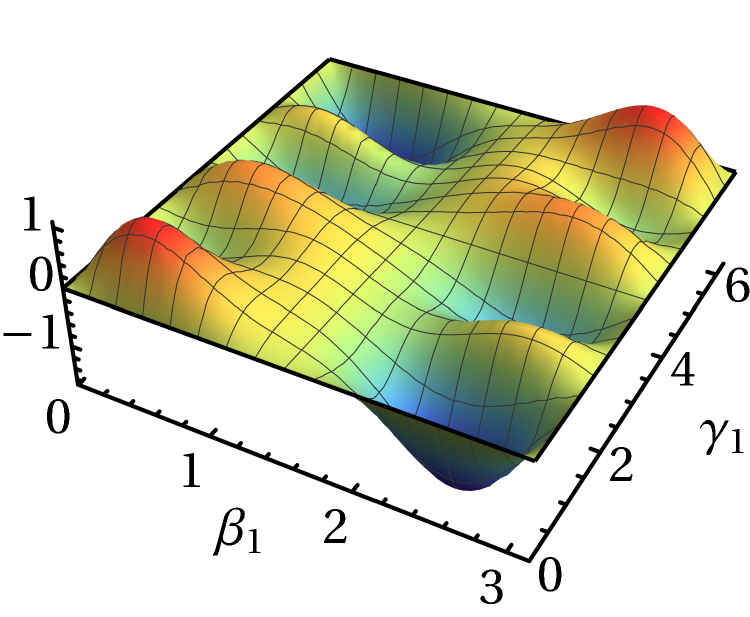}
    &
    \raisebox{12em}{(b)}
    &
    \includegraphics[width=0.4\textwidth]{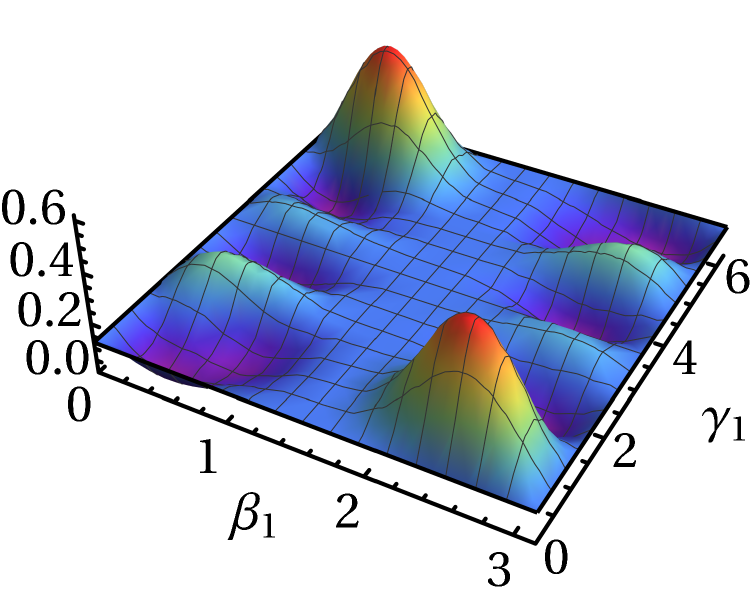}
\end{tabular}
 \caption{Example QAOA result for $p=1$ that shows (a) the expectation value of the energy $\langle E\rangle$ (see Eq.~(\ref{eq:qaoaenergyexpectation})) and (b) the probability to measure the solution state $\vert\braket{010|\psi}\vert^2$. In this case, the energy minimum at $(\beta_1,\gamma_1)\approx(2.505,0.681)$ is very close to the point $(\beta_1,\gamma_1)\approx(2.524,0.713)$ where the success probability is maximal. The state vector $\ket\psi$ has been obtained by simulating the QAOA circuit in Fig.~\ref{fig_michielsen_qaoa_example} for the whole range of $\beta_1\in[0,\pi)$ and $\gamma_1\in[0,2\pi)$. Beyond this range, the landscapes can be continued periodically; the periodicity in $\beta_1$ and $\gamma_1$ is due to the periodicity of the $R^x$ and $R^z$ gates in Fig.~\ref{fig_michielsen_qaoa_example}.}
\label{fig_michielsen_qaoa_result}
\end{figure}

Constructing the QAOA circuit for $p=1$ according to Figs.~\ref{fig_michielsen_qaoa} and \ref{fig_michielsen_zz_identity} yields the quantum circuit in Fig.~\ref{fig_michielsen_qaoa_example}. The program to simulate this circuit for a certain range of values for $\beta_1$ and $\gamma_1$ is shown in Listing~\ref{juniqexampleqaoa}. 

Figure~\ref{fig_michielsen_qaoa_result} shows the expectation value of the energy, computed from the state vector $\ket\psi$ according to Eq.~(\ref{eq:qaoaenergyexpectation}), and the success probability $\vert\braket{010|\psi}\vert^2$. In this case, the energy minimum (see Fig.~\ref{fig_michielsen_qaoa_result}(a)) is very close to the point with maximum success probability (see Fig.~\ref{fig_michielsen_qaoa_result}(b)). Note that this is not guaranteed for larger QAOA applications \cite{Willsch2021JUQCSGQAOA,Willsch2019BenchmarkingQAOA}.
\\~\\
\textbf{Exercise 5:} Run the $p=1$ QAOA for the energy function in Eq.~(\ref{eq:qaoaenergyexample}) by simulating the circuit in Fig.~\ref{fig_michielsen_qaoa_example}. Either perform a grid scan over $\beta_1\in[0,\pi)$ and $\gamma_1\in[0,2\pi)$ (as done to produce Fig.~\ref{fig_michielsen_qaoa_result}), or write an optimization program to find the location of the minimum (e.g.~by using the \texttt{scipy} package\cite{scipy}).
\\~\\
\textbf{Exercise 6:} Construct the $p=2$ QAOA circuit and run it to improve upon the $p=1$ result. One way to do this is to take the best $(\beta_1,\gamma_1)$ from Exercise 5 or Fig.~\ref{fig_michielsen_qaoa_result}, and optimize for $(\beta_2,\gamma_2)$. Another way would be to consider a case with larger $p$ and take values for $(\beta_k,\gamma_k)$ from a quantum annealing schedule as described in \cite{Willsch2021JUQCSGQAOA}.

\begin{lstlisting}[float, language=Python, caption={Example program for the $p=1$ QAOA circuit shown in Fig.~\ref{fig_michielsen_qaoa_example}. Separate circuits are simulated for all given values of $\beta_1$ and $\gamma_1$. From the resulting state vector, both the energy and the success probability are evaluated. The best result in the output should be at $\beta_1=2.5$ and $\gamma_1=0.7$ with energy $-1.69$ and success probability $58.9\%$.}, label=juniqexampleqaoa]
import qiskit

h = [-1, 1/2, -1/2]
J = [1/2, 1/2]
beta1s = [2.0, 2.5]
gamma1s = [0.5, 0.7, 1.0]

def E(s0, s1, s2):
    return h[0]*s0 + h[1]*s1 + h[2]*s2 + J[0]*s0*s1 + J[1]*s1*s2

qaoa_circuits = []
qaoa_parameters = []

for beta1 in beta1s:
    for gamma1 in gamma1s:
        circuit = qiskit.QuantumCircuit(3)
        circuit.h(0)
        circuit.h(1)
        circuit.h(2)
        circuit.rz(2*gamma1*h[0], 0)
        circuit.rz(2*gamma1*h[1], 1)
        circuit.rz(2*gamma1*h[2], 2)
        circuit.cx(0,1)
        circuit.rz(2*gamma1*J[0], 1)
        circuit.cx(0,1)
        circuit.cx(1,2)
        circuit.rz(2*gamma1*J[1], 2)
        circuit.cx(1,2)
        circuit.rx(2*beta1, 0)
        circuit.rx(2*beta1, 1)
        circuit.rx(2*beta1, 2)
        qaoa_circuits.append(circuit.reverse_bits())
        qaoa_parameters.append([beta1, gamma1])

from juqcs import Juqcs
backend = Juqcs.get_backend('statevector_simulator')
backend.allocate(minutes=30, max_qubits=3)

job = qiskit.execute(qaoa_circuits, backend=backend)
result = job.result()

print('beta1\tgamma1\tenergyexpectation\tsuccessprobability')
for i,(beta1,gamma1) in enumerate(qaoa_parameters):
    psi = result.get_statevector(i)
    prob = abs(psi)**2
    energy  = E(+1,+1,+1)*prob[0b000] + E(+1,+1,-1)*prob[0b001]
    energy += E(+1,-1,+1)*prob[0b010] + E(+1,-1,-1)*prob[0b011]
    energy += E(-1,+1,+1)*prob[0b100] + E(-1,+1,-1)*prob[0b101]
    energy += E(-1,-1,+1)*prob[0b110] + E(-1,-1,-1)*prob[0b111]
    probability = prob[0b010]
    print(f'{beta1}\t{gamma1}\t{energy}\t{probability}')

backend.deallocate()
\end{lstlisting}

\clearpage

\section{Quantum annealing}
The second part of these lecture notes focuses on quantum annealing\index{quantum annealing}. Besides gate-based quantum computers, quantum annealing has emerged as the second major paradigm of quantum computing. As quantum annealers are somewhat simpler to manufacture, much larger devices of 5000+ qubits are already available and the technology is closer to the verge of technological maturity.

For this reason, the quantum annealing programs discussed in Section~\ref{sec:programming_dwave} of these lecture notes exclusively target real devices, in contrast to the simulators discussed in the gate-based case above. Furthermore, the program type reflects the typical kind of problems solved on current quantum annealers that are a little bit closer to actual real-world applications \cite{Willsch2019BenchmarkingQAOA,Willsch2020QSVM,calaza2021gardenoptimization,Willsch2021BenchmarkAdvantage}. We will take a particular look at the garden optimization problem \cite{calaza2021gardenoptimization} as an application.

\subsection{Optimization problems with binary variables}
Optimization of parameters in high-dimensional spaces can in general  be a (computationally) demanding task. 
Gradient-based algorithms as well as non-gradient based algorithms usually require many evaluations of the cost function or its gradient. Typically, they rely on the continuity of the parameter space and often also on the continuity of the function itself.
In high-dimensional spaces, cost functions usually have many local optima in which optimization algorithms can get stuck, as well as plateaus which slow down the convergence.
There exist algorithms that use an adaptive step size to mitigate these effects.
However, usually it is still necessary to start the optimizer with different (random) initial parameters and take the best solution that was returned at the end.
In that sense, these algorithms also do not guarantee that the globally optimal solution has been found.

For cost functions of discrete or binary variables, these optimization algorithms are not applicable because the functions are only defined at discrete values.
In general such optimization problems are NP-hard, and only a brute-force search over all possible inputs can deliver the optimal solution.
However, this is infeasible for large configuration spaces.
For many discrete optimization\index{discrete optimization} problems, heuristic algorithms have been developed which work well for many cases.
A heuristic algorithm may find the globally optimal solution, but it may also return just a locally optimal solution as the outcome usually depends on the initialization.
The advantage of heuristic algorithms is that the run time is much shorter than for a brute-force search.
An example for such a heuristic algorithm is the simulated annealing algorithm~\cite{kirkpatrick83}.

A common discrete optimization problem in physics is to find the ground state of the Ising spin Hamiltonian:
\begin{equation}
    \label{eq:ising}
    H_\mathrm{Ising} = \sum\limits_{i=1}^Nh_i\sigma^z_i + \sum\limits_{i<j}J_{ij}\sigma^z_i\sigma^z_j,
\end{equation}
where $N$ denotes the number of spins, $h_i$ denotes the local field for spin $i$, $J_{ij}$ denotes the coupling strength between spins $i$ and $j$, and $\sigma^z_i$ denotes the Pauli $z$ matrix
\begin{equation}
    \sigma^z = \begin{pmatrix} 1 & 0 \\ 0 & -1 \end{pmatrix}
\end{equation}
for spin $i$ with eigenstates $\ket{\uparrow}$ and $\ket{\downarrow}$ such that $\sigma^z\ket{\uparrow}=\ket{\uparrow}$ and $\sigma^z\ket{\downarrow}=-\ket{\downarrow}$.
This kind of problems can be solved on the D-Wave quantum annealer.

Another form of discrete optimization that can be solved on the D-Wave quantum annealer is Quadratic Unconstrained Binary Optimization (QUBO\index{QUBO}). The QUBO cost function to minimize is given in the form
\begin{equation}
    C(\mathbf{x}) = \sum\limits_{i=1}^Na_i x_i + \sum\limits_{i<j}b_{ij}x_ix_j,
\end{equation}
where $N$ is the number of binary variables $x_i\in\{0,1\}$, $\mathbf{x}=x_1x_2\dots x_N$ denotes the bitstring obtained by concatenating the problem variables, $a_ix_i$ are the linear terms and $b_{ij}x_ix_j$ are the quadratic terms in the binary variables, and $a_i, b_{ij}\in \mathbb{R}$ are the parameters that define the problem instance to be solved. 
In the definition of a QUBO, the term ``Unconstrained'' means that there are no constraints in the optimization of $C(\mathbf{x})$ such as ``subject to $f(\mathbf{x})=0$".
Solving a QUBO is equivalent to solving for the ground state of an Ising Hamiltonian Eq.~(\ref{eq:ising}).
Many optimization problems that can be formulated in terms of an Ising or QUBO model are discussed in \cite{lucas14}.

\subsection{Working principle of a quantum annealer}
Initially, quantum annealing\index{quantum annealing} was invented as a heuristic algorithm for a classical computer~\cite{Apolloni89,finnila94,kadowaki98}.
It was inspired by the simulated annealing algorithm~\cite{kirkpatrick83} where thermal fluctuations are replaced by quantum fluctuations. For simulated annealing, thin but high energy barriers are difficult to overcome as thermal hopping processes become unlikely for high barriers. Therefore, the idea was that quantum fluctuations in quantum annealing allow for tunneling through these thin but high energy barriers.
The quantum annealing Hamiltonian can be expressed as
\begin{align}
    H_\mathrm{QA}(t/t_\mathrm{max}) =\Gamma(t/t_\mathrm{max})H_\mathrm{QF} + H_\mathrm{problem} ,
    \label{eq:H_QA}
\end{align}
where $H_\mathrm{problem}$ encodes the optimization problem to be solved, and $H_\mathrm{QF}$ denotes the Hamiltonian introducing the quantum fluctuations.
The function $\Gamma(t/t_\mathrm{max})$ controls the strength of these fluctuations. It has to satisfy $\Gamma(0)\gg$ energy scale of $H_\mathrm{problem}$ and $\Gamma(1)\approx 0$.

\begin{figure}[t!]
 \centering
 \includegraphics[width=0.7\textwidth]{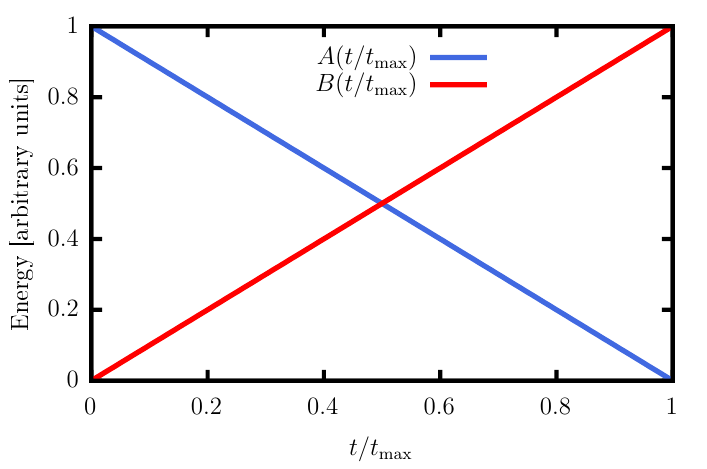}
 \caption{The annealing functions $A(t/t_\mathrm{max})$ and $B(t/t_\mathrm{max})$ for a linear annealing schedule.}
 \label{fig_annealing_scheme_lin}
\end{figure}

Later, adiabatic quantum computation\index{adiabatic quantum computation} was proposed~\cite{farhi00,childs01}.
The idea to perform the computation is based on the adiabatic theorem: The quantum system is to be initialized in the known ground state of an initial Hamiltonian $H_\mathrm{init}$.
Then, the time-dependent Hamiltonian of the system
\begin{equation}
    H(t/t_\mathrm{max}) = A(t/t_\mathrm{max})H_\mathrm{init} + B(t/t_\mathrm{max})H_\mathrm{final}
    \label{eq:H_aqc}
\end{equation}
is changed over time into the Hamiltonian $H_\mathrm{final}$ which encodes an optimization problem (e.g., the Ising Hamiltonian Eq.~(\ref{eq:ising})). The annealing functions $A(t/t_\mathrm{max})$ and $B(t/t_\mathrm{max})$ satisfy $A(0) \gg B(0)$ and $B(1)\gg A(1)$. An example for a linear annealing schedule is shown in Fig.~\ref{fig_annealing_scheme_lin}.
With $H_\mathrm{final}=H_\mathrm{problem}$, $H_\mathrm{init}=H_\mathrm{QF}$, $A(t/t_\mathrm{max})=\Gamma(t/t_\mathrm{max})$ and $B(t/t_\mathrm{max})=1$, the Hamiltonian Eq.~(\ref{eq:H_aqc}) implements the quantum annealing Hamiltonian Eq.~(\ref{eq:H_QA}).
The adiabatic theorem~\cite{born28} states that if the Hamiltonian $H(t/t_\mathrm{max})$ changes sufficiently slowly with time, the system stays in the instantaneous ground state so that a measurement at the end of the process yields the state that encodes the solution to the optimization problem.
A detailed review on the ideal (i.e., closed-system) model of adiabatic quantum computation is given in \cite{albash18}.

For a single spin-1/2 particle in a time-dependent external magnetic field, the closed-system model can be solved analytically exactly and is described by the Landau-Zener theory~\cite{zener32}:

The Hamiltonian describing the spin in the external field $h_z=vt$ which changes with time $t$ from $-\infty$ to $\infty$ with sweep velocity $v$ and a time-independent transverse field $h_x$ is given by
\begin{align}
  H_\mathrm{LZ}(t) = -h_x\sigma^x -h_z(t)\sigma^z = -h_x\sigma^x -vt\sigma^z
\end{align}
where $\sigma^x$ is the Pauli $x$ matrix
\begin{equation}
    \sigma^x = \begin{pmatrix} 0 & 1 \\ 1 & 0 \end{pmatrix},
\end{equation}
with eigenstates $\ket{+} = (\ket{\uparrow}+\ket{\downarrow})/\sqrt{2}$ and $\ket{-}=(\ket{\uparrow}-\ket{\downarrow})/\sqrt{2}$.
The spin is prepared in the ground state of $H_\mathrm{LZ}(t\to-\infty)$ which is the state $\ket{\psi_\mathrm{init}}=\ket{\downarrow}$.
The spin then evolves with the Hamiltonian $H_\mathrm{LZ}(t)$ and the Landau-Zener theory gives the probability to find the spin in the ground ($\ket{\uparrow}$) or excited state ($\ket{\downarrow}$) of $H_\mathrm{LZ}(t)$ for $t\to\infty$:
\begin{align}
    P_\uparrow=1-e^{-\pi h_x^2/v}, \quad P_\downarrow = e^{-\pi h_x^2/v}.
\end{align}
These probabilities show that for fast sweeping (large $v$), $P_\uparrow\to 0$ and $P_\downarrow\to 1$. They also show that for $h_x\to 0$, $P_\uparrow\to 0$ and $P_\downarrow\to 1$. Since the minimal energy gap is proportional to $|h_x|$, the probabilities scale not only with the sweep velocity but also with the minimal energy gap squared.

In practice however, the annealing system is always connected to an environment at a finite temperature. This means that some sources of noise can never be removed completely. So not only might a too rapid change of the Hamiltonian excite the system but too slow of an annealing procedure might as well.

\subsection{Architecture of D-Wave quantum annealers}
\begin{figure}[t!]
 \centering
 \includegraphics[width=0.9\textwidth]{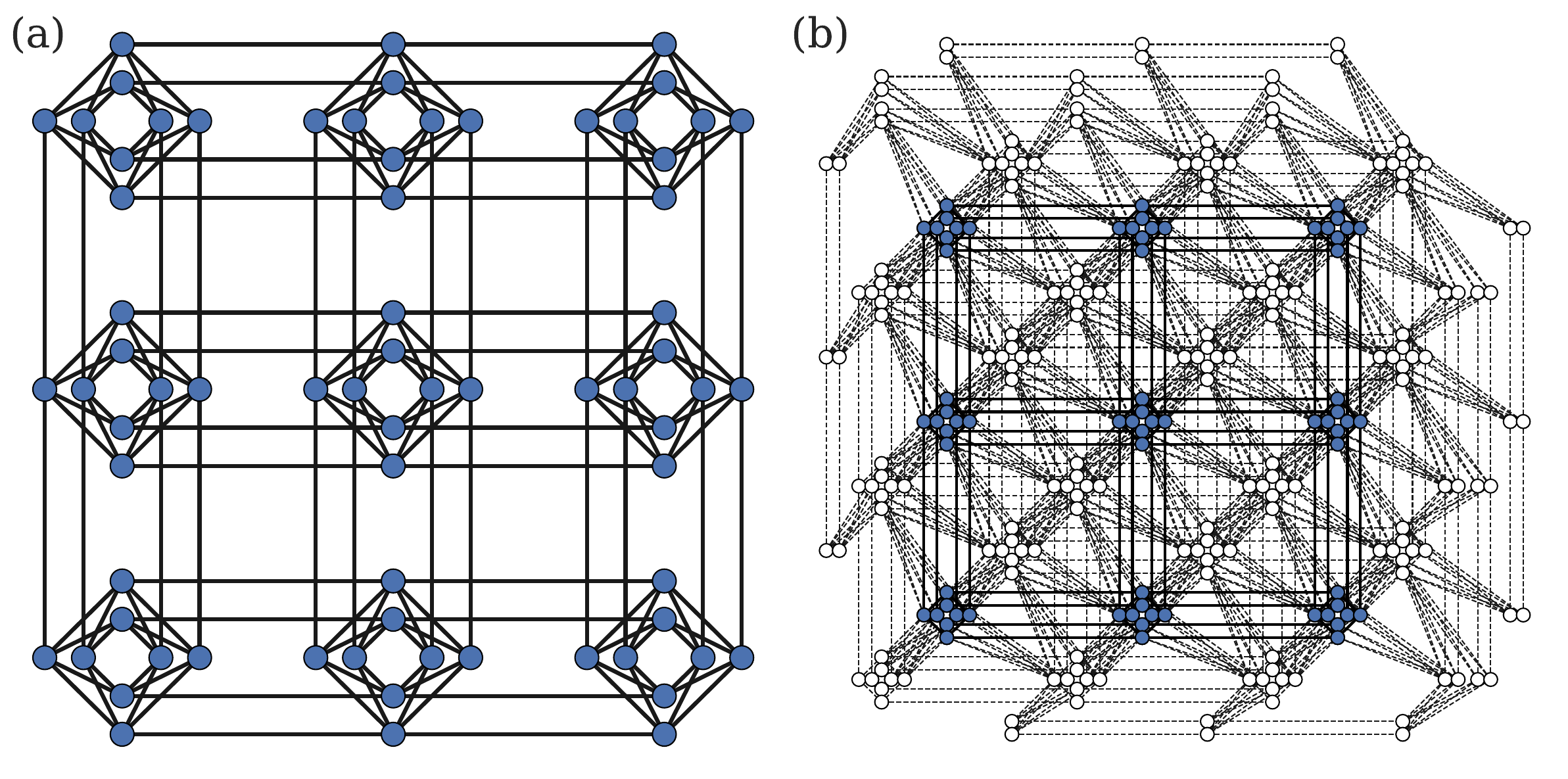}
 \caption{Small-size examples of the connectivity graphs of the (a) Chimera topology of the D-Wave 2000Q processors and (b) Pegasus topology of the Advantage processors. Blue nodes in the Pegasus graph show the embedding of the Chimera graph onto the Pegasus graph.}
 \label{fig:chimera_pegasus}
\end{figure}

There are currently two generations of D-Wave quantum annealers available: The previous generation (D-Wave 2000Q) with about 2000 qubits and up to 6 couplers per qubit, and the current generation (Advantage) with more than 5000 qubits and up to 15 couplers per qubit.
The qubits of the D-Wave 2000Q quantum processors are arranged in the so-called Chimera topology and the qubits of the Advantage systems are arranged in the so-called Pegasus topology (see Fig.~\ref{fig:chimera_pegasus}). The Chimera graph is a subgraph of the Pegasus graph.

\begin{figure}[t!]
 \centering
 \includegraphics[width=0.7\textwidth]{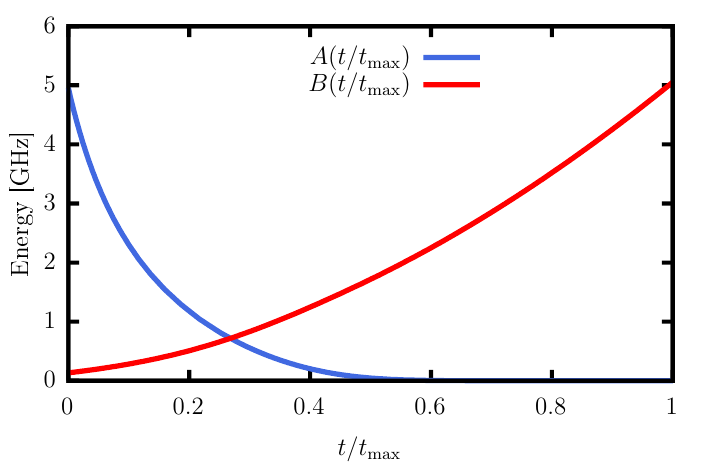}
 \caption{The annealing functions $A(t/t_\mathrm{max})$ and $B(t/t_\mathrm{max})$ of the annealing schedule of D-Wave's 5000+ qubit processor \texttt{Advantage\_system1.1}.}
 \label{fig_annealing_scheme_dwave}
\end{figure}

In the case of D-Wave quantum annealers, the final Hamiltonian is given by the Ising Hamiltonian
\begin{equation}
    H_\mathrm{final} = H_\mathrm{Ising} = \sum\limits_{i=1}^Nh_i\sigma^z_i +\!\!\! \sum\limits_{\substack{i<j\\i,j\text{ neighbors}}}\!\!\!J_{ij}\sigma^z_i\sigma^z_j,
    \label{eq:H_final}
\end{equation}
and the initial Hamiltonian is given by
\begin{equation}
    H_\mathrm{init} = -\sum\limits_{i=1}^N\sigma^x_i.
    \label{eq:H_init}
\end{equation}
The ground state of the initial Hamiltonian is given by the equal superposition of all computational basis states,
\begin{align}
    \ket{\psi_\mathrm{init}} = \ket{+}^{\otimes N} = \frac{1}{2^{N/2}}\sum\limits_{z_i\in\{\uparrow,\downarrow\}} \ket{z_1 z_2 \dots z_N}.
\end{align}
The qubits (two-level systems) of the Hamiltonians in Eqs.~(\ref{eq:H_final}) and (\ref{eq:H_init}) are built of superconducting circuits.
The particular design of these circuits is called flux qubits.
The flux qubits as well as the couplers which allow for a tunable coupling between the qubits are controlled via external fluxes.
The parameters $h_i$ and $J_{ij}$ of the final Hamiltonian given in Eq.~(\ref{eq:H_final}) are controlled by time-independent external magnetic fluxes.
The annealing process is controlled through time-dependent external magnetic fluxes which change the effective Hamiltonian from the initial Hamiltonian to the final Hamiltonian.
As an example, in Fig.~\ref{fig_annealing_scheme_dwave} we show the annealing schedule of the processor \texttt{Advantage\_system1.1}.
The annealing schedule is obviously different from the linear annealing schedule shown in Fig.~\ref{fig_annealing_scheme_lin}.
The reason for this is that the hardware design of the flux qubits does not allow for an independent control of the functions $A(t/t_\mathrm{max})$ and $B(t/t_\mathrm{max})$~\cite{harris10_eightqubit}.
Like all superconducting qubits, these systems are actually multi-level systems of which only a two-dimensional subspace (spanned by the two lowest energy eigenstates) functions as the qubit.
More detailed information can be found in, for instance, Refs.~\cite{harris10_eightqubit,harris09,harris10}.

\subsection{Limitations}
Current quantum annealers are subject to various limitations which may have an impact on the performance.
For instance, due to the limited connectivity of the qubits (up to 6 connections per qubit on the Chimera architecture; up to 15 connections per qubit on the Pegasus architecture), the variables of a given problem may not be directly mappable to the qubits. In that case, an \emph{embedding} of the problem graph onto the hardware graph is necessary (see Section \ref{sec:embedding} for further details).
Finding an (optimal) embedding is in general a hard problem itself.
However, there exist heuristic methods to find embeddings and the \texttt{Ocean} package provides such an algorithm (cf.~Section~\ref{sec:embedding}).
The need to find an embedding limits the performance due to the time overhead required to generate an embedding and it often also increases the required number of qubits (see Section~\ref{sec:embedding} for more details).
Thus, although problem instances that feature the native hardware graphs (Chimera or Pegasus graphs) can be put on the D-Wave 2000Q and Advantage systems with sizes of 2000+ and 5000+ qubits, respectively, only 64- and 124-qubit fully-connected problems can be placed on these systems, respectively.

Other limiting factors are the restricted range and the limited precision of the parameters $h_i$ and $J_{ij}$. In order for the parameter values to fit into the available range they have to be rescaled (see Section~\ref{sec:specification}). Thus, for problem instances which cover a large range of values but at the same time include parameters with small differences, these differences may not be resolvable anymore within the available precision, i.e., configured parameters on the hardware may differ from the specified ones.

A general limiting factor in practical quantum annealing is that the quantum system can never be completely isolated from the environment and noise sources. As a consequence, the system can be thermally excited. In particular, for theoretical closed-system quantum annealing, the annealing time cannot be too long. However, for open systems, a long annealing time means the system has a longer time during which it can interact with the environment, leading to a higher probability of leaving the ground state.

Apart from these practical limitations, there are also theoretical issues.
The adiabatic theorem states that the quantum system stays in its instantaneous ground state if the Hamiltonian changes sufficiently slowly in time. In particular, the annealing time should satisfy ($s=t/t_\mathrm{max}$)~\cite{amin09_adiabatictheorem}
\begin{equation}
        t_\mathrm{max} \gg \max_{s\in[0,1]} \frac{\left| \bra{E_n(s)} \frac{\partial H}{\partial s} \ket{E_0(s)} \right|}{(E_0(s)-E_n(s))^2} \quad\text{ for } n\neq 0.
\end{equation}
Thus, for an adiabatic evolution and for exponentially small energy gaps, the annealing time is also expected to increase exponentially.

\subsection{Programming a D-Wave quantum annealer}\label{sec:programming_dwave}
In this section, we focus on programming the D-Wave quantum annealer using D-Wave's Python package \texttt{Ocean-SDK}~\cite{dwave_github,dwave_documentation} as well as some practical aspects which are useful when programming a D-Wave quantum annealer. It is worth mentioning that, even though the previous sections contain useful technical details for background information, the only knowledge required to program a quantum annealer is about the kind of optimization problem (Ising or QUBO) that it solves. This makes quantum annealers particularly attractive for non-physicists.

\subsubsection{Problem specification}\label{sec:specification}
Optimization problems that can be put onto D-Wave quantum annealers have to be formulated as QUBO
\begin{equation}
    \min_{x_i=0,1} \left(\sum\limits_{i=0}^{N-1}a_i x_i + \sum\limits_{i<j}^{N-1}b_{ij}x_ix_j\right) = \min_{x_i=0,1} \left( \sum\limits_{i\le j}^{N-1}x_iQ_{ij}x_j\right),
\end{equation}
or as an Ising Hamiltonian
\begin{equation}
  \min_{s_i=\pm 1} \left( \sum\limits_{i=0}^{N-1}h_i s_i + \sum\limits_{i<j}^{N-1}J_{ij}s_is_j \right),
  \label{eq:opt_ising}
\end{equation}
where $s_i\in\{-1,+1\}$ are the eigenvalues of the Pauli-$z$ matrix $\sigma_i^z$.
Reformulating a QUBO as Ising Hamiltonian can be done by substituting
\begin{align}
  x_i = \frac{1+s_i}{2}.
  \label{eq:substitution}
\end{align}
This convention is commonly used in quantum annealing and maps $x_i=0$ and $x_i=1$ to $s_i=-1$ and $s_i=1$, respectively. Note that this convention is different from the one commonly used for gate-based quantum computing (see above).
When converting problems between QUBO and Ising formulation, constants which arise from the substitution Eq.~(\ref{eq:substitution}) can be neglected as they only lead to an energy shift but do not change the solution of the optimization.

To specify the problem instance using the \texttt{Ocean-SDK}, the coefficients can be stored in a dictionary \texttt{Q = \{($i$,$j$):$Q_{ij}$\}} (QUBO) or two dictionaries \texttt{h = \{$i$:$h_i$\}} and \texttt{J = \{($i$,$j$):$J_{ij}$\}} (Ising).

In practice, the coupling strength $J_{ij}$ can only be set to values in a certain interval $[J_\mathrm{min},J_\mathrm{max}]$, where usually $J_\mathrm{min}=-J_\mathrm{max}$, with a limited precision. The same applies to the single-qubit bias $h_i$ with interval $[h_\mathrm{min},h_\mathrm{max}]$.
Thus, all $h_i$ and $J_{ij}$ need to be rescaled by a factor of~\cite{DWaveSolversParameters}
\begin{equation}
  r=\max\left\{
  \max\!\left[\frac{\max\{h_i\}}{h_{\mathrm{max}}},0\right]\!,
  \max\!\left[\frac{\min\{h_i\}}{h_{\mathrm{min}}},0\right]\!,
  \max\!\left[\frac{\max\{J_{ij}\}}{J_{\mathrm{max}}},0\right]\!,
  \max\!\left[\frac{\min\{J_{ij}\}}{J_{\mathrm{min}}},0\right]
  \right\},
  \label{eq:rescale}
\end{equation}
to fit into the ranges $h_\mathrm{min} \le h_i \le h_\mathrm{max}$ and $J_\mathrm{min} \le J_{ij} \le J_\mathrm{max}$.
If $r<1$, rescaling is optional, but it may be useful to exhaust a larger parameter range and potentially improve the performance.
When \texttt{auto\_scale} is set to \texttt{true} (default), the rescaling is  performed automatically when submitting a problem through \texttt{Ocean}. When auto-scaling is disabled and the problem parameters do not fit into the parameter range, the submission of the problem fails.
Note that rescaling does not change the solution of the problem.
In general, it is advised to have the auto-scaling feature enabled (default).
However, there may be certain cases where it has to be disabled.

\subsubsection{Submitting a problem to the D-Wave quantum annealer through the Ocean package}\label{sec:programming_dwave_details} 

\begin{lstlisting}[float, language=Python, caption=A minimal working example to run a program on a D-Wave quantum annealer., label=example1]
from dwave.system import DWaveSampler

sampler = DWaveSampler(solver='DW_2000Q_6', token='insert_your_token_here')
h = {0:1, 4:-0.5}
J = {(0,4):1, (0,5):-1}

response = sampler.sample_ising(h, J)
print(response)
\end{lstlisting}

Listing~\ref{example1} shows a minimal working example of a submission to the D-Wave quantum annealer.
The class \texttt{DWaveSampler} takes a solver (for example the, at the time of writing, current hardware solvers `\texttt{DW\_2000Q\_6}' or `\texttt{Advantage\_system1.1}') to which to submit the problem.
If one did not create a \texttt{config}-file during or after the installation of the \texttt{Ocean-SDK}, the personal token also has to be provided to \texttt{DWaveSampler}.
Depending on the formulation of the problem (Ising or QUBO), one has to create the \texttt{h} and \texttt{J} or \texttt{Q} dictionaries, respectively.
The example code in Listing~\ref{example1} shows the case for the Ising formulation.
For larger problems, the dictionaries should be generated algorithmically.
The class \texttt{DWaveSampler} has the member functions \texttt{sample\_ising} and \texttt{sample\_qubo} which submit the specified problem to the QPU.
The function \texttt{sample\_ising} takes the \texttt{h} and \texttt{J} dictionaries and the function \texttt{sample\_qubo} takes the \texttt{Q} dictionary.

Optional parameters of the functions \texttt{sample\_ising} and \texttt{sample\_qubo} are for instance the number of samples obtained per submission (\texttt{num\_reads}), the \texttt{annealing\_time} (in \textmu s), or the \texttt{chain\_strength} (see Section~\ref{sec:embedding}). An example is shown in Listing~\ref{example2}.

\begin{lstlisting}[float, language=Python, caption=Example showing how to use an \texttt{EmbeddingComposite}., label=example2]
from dwave.system import DWaveSampler, EmbeddingComposite

sampler = EmbeddingComposite(DWaveSampler(solver='DW_2000Q_6', token='insert_your_token_here'))
Q = {(0,0):1, (0,1):1, (0,2):-1, (1,2):-0.8}

response = sampler.sample_qubo(Q, num_reads=100, chain_strength=2, annealing_time=5)
print(response)
\end{lstlisting}

Listing~\ref{example2} also illustrates how to use an \texttt{EmbeddingComposite} (see Section~\ref{sec:embedding}).
The return value contains the solutions returned by the quantum annealer as well as some additional information.
An example output is shown in Listing~\ref{example_output}:
The first column labels the different solutions returned (rows).
The following $N_\mathrm{qubit}$ (in this case $N_\mathrm{qubit}=3$) columns give the values of the qubits in the returned solution.
The next three columns contain the energy, the number of occurrences of this solution in all samples, and the chain break fraction (see Section~\ref{sec:embedding}), respectively.
In our example output, we obtained 79 times the solution no.~0 with energy $-0.8$ (the energy of the ground state) and no chain breaks.
In addition, the \texttt{response} also contains the information that the result is given in \texttt{BINARY}, i.e.~QUBO, format (in the Ising representation it would be \texttt{SPIN}), that 7 distinct answers were returned by the quantum annealer (the number of rows), that the number of samples is 100 (equals \texttt{num\_reads}), and that the number of qubits (variables) is three.
Further information can be accessed through \texttt{response.info}.

\begin{lstlisting}[float, caption=Example output of the program given in Listing~\ref{example2}., label=example_output, style=plain]
   0  1  2 energy num_oc. chain_.
0  0  1  1   -0.8      79     0.0
1  0  0  1    0.0       5     0.0
2  1  0  1    0.0       5     0.0
3  0  0  0    0.0       7     0.0
4  0  1  0    0.0       1     0.0
5  1  1  1    0.2       2     0.0
6  1  0  0    1.0       1     0.0
['BINARY', 7 rows, 100 samples, 3 variables]
\end{lstlisting}

\begin{figure}[t!!]
 \centering
 \includegraphics[width=1.0\textwidth]{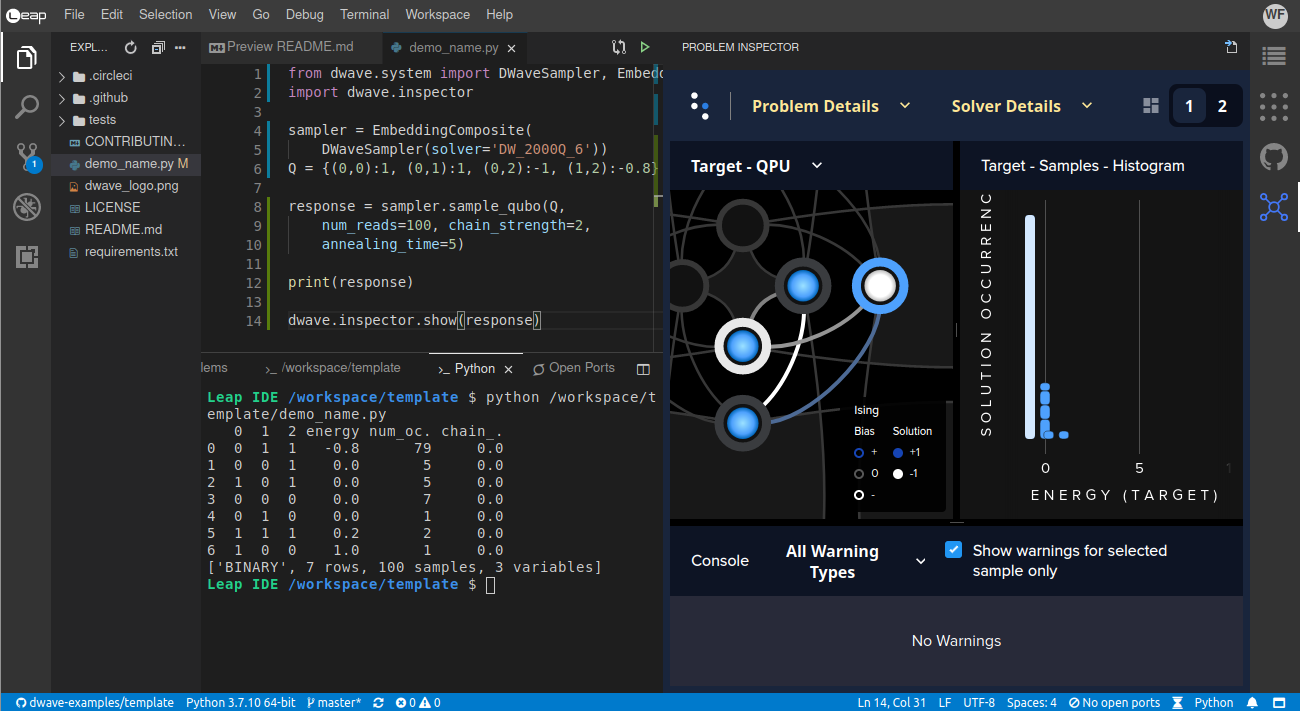}
 \caption{Running the example program from Listing~\ref{example2} and visualizing the result on the D-Wave using the inspector (see Listings~\ref{example_output} and \ref{example_inspector}). The Leap IDE with an example template to run such an experiment can be opened at \protect\url{https://ide.dwavesys.io/\#https://github.com/dwave-examples/template}.}
 \label{fig_michielsen_inspector}
\end{figure}

Another tool of the \texttt{Ocean} package which can be handy when studying the returned results is \texttt{dwave.inspector}. With this tool, the response object can be visualized. The usage is illustrated in Listing~\ref{example_inspector} and Fig.~\ref{fig_michielsen_inspector}.
\begin{lstlisting}[float, language=Python, caption={Example showing how to use \texttt{dwave.inspector} to visualize the result. A screenshot using the Leap IDE is shown in Fig.~\ref{fig_michielsen_inspector}.}, label=example_inspector]
import dwave.inspector

##################
#  previous code #
##################

dwave.inspector.show(response)
\end{lstlisting}

\subsubsection{Implementing constraints}\label{sec:constraints}
As mentioned previously, D-Wave quantum annealers are desgined to solve QUBO or Ising problems by minimizing the corresponding energy function without constraints.
However, in practice optimization problems often require constrained optimization, i.e., the objective function $C(\mathbf{x})$ needs to be minimized under a certain constraint $f(\mathbf{x})=c$, where $f$ is a function that takes a string of bits $\mathbf{x}$ and returns a scalar, and $c$ is a scalar, i.e., the optimization task is to ``minimize $C(\mathbf{x})$ subject to $f(\mathbf{x})=c$".

We can consider such a constraint by formulating it as a QUBO (or Ising Hamiltonian) and adding it to the objective function:
We can write the constraint $f(\mathbf{x})=c$ as $f(\mathbf{x})-c=0$.
The square $(f(\mathbf{x})-c)^2$ is always greater than zero, except if the constraint $f(\mathbf{x})=c$ is satisfied in which case the square is equal to zero and thus minimal. By adding the square term $(f(\mathbf{x})-c)^2$ to the objective function, we effectively add a penalty to the objective function if the constraint is not satisfied.
To keep things simple, we assume that $f(\mathbf{x})$ is a linear function in the bits $x_i$ (further information can be found in \cite{dwave_constraints}).
We add the penalty term to the objective function $C(\mathbf{x})$ so that the new/modified objective function reads $C(\mathbf{x})+\lambda (f(\mathbf{x})-c)^2$, where $\lambda$ is a scalar called Lagrange multiplier and has to be chosen reasonably.
The optimization task is now ``minimize $C(\mathbf{x})+\lambda (f(\mathbf{x})-c)^2$".

A ``reasonable" choice for $\lambda$ means that $\lambda$ should neither be too small nor too large.
If $\lambda$ is chosen too small, the constraint will likely not be satisfied in the optimal solution for $C(\mathbf{x})+\lambda (f(\mathbf{x})-c)^2$ as it may be more favourable to accept a penalty multiplied by a small $\lambda$ than to return a state with a larger cost function value $C(\mathbf{x})$.
On the other hand, a too large $\lambda$ will force the constraint to be satisfied but due to the rescaling of the parameters (see Section~\ref{sec:specification}), the problem parameters might become so small that they cannot be represented accurately enough with the limited precision, and also the energy differences become so small that excitations to higher energy states, which do not encode the optimal solution to the original problem anymore, become very likely.

\subsubsection{Embedding of problem graphs onto the hardware graph}\label{sec:embedding}
\begin{figure}[t!]
 \centering
 \includegraphics[width=0.35\textwidth]{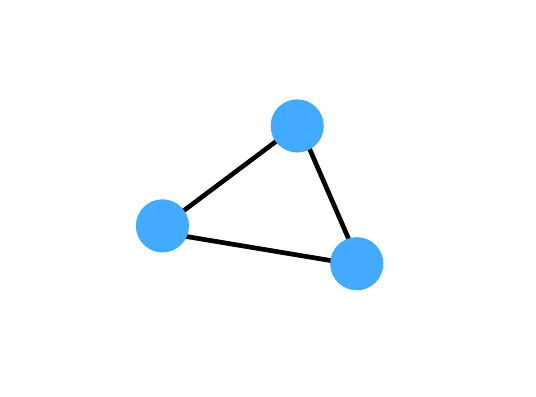}
 \includegraphics[width=0.6\textwidth]{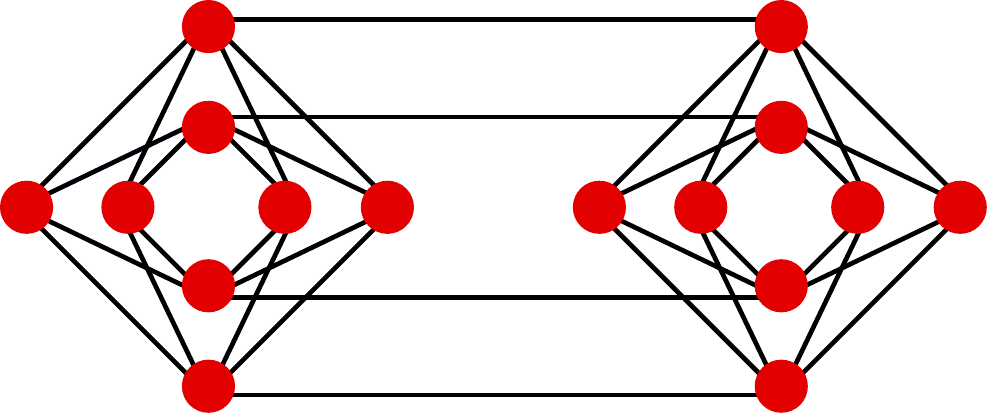}
 \caption{A graph with triangular connectivity (left) cannot be directly mapped onto the Chimera hardware graph (right).}
 \label{fig_triangular_connectivity}
\end{figure}

\begin{figure}[t!]
 \centering
 \includegraphics[width=0.35\textwidth]{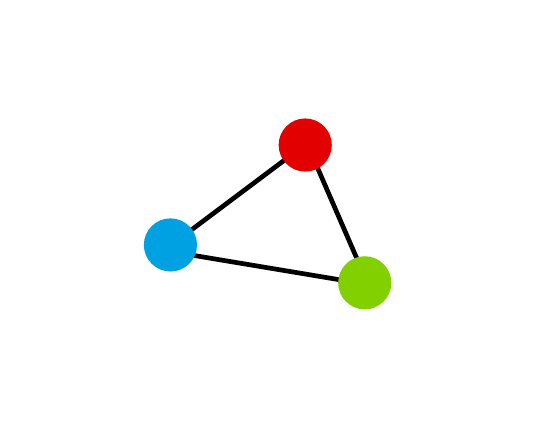}
 \includegraphics[width=0.6\textwidth]{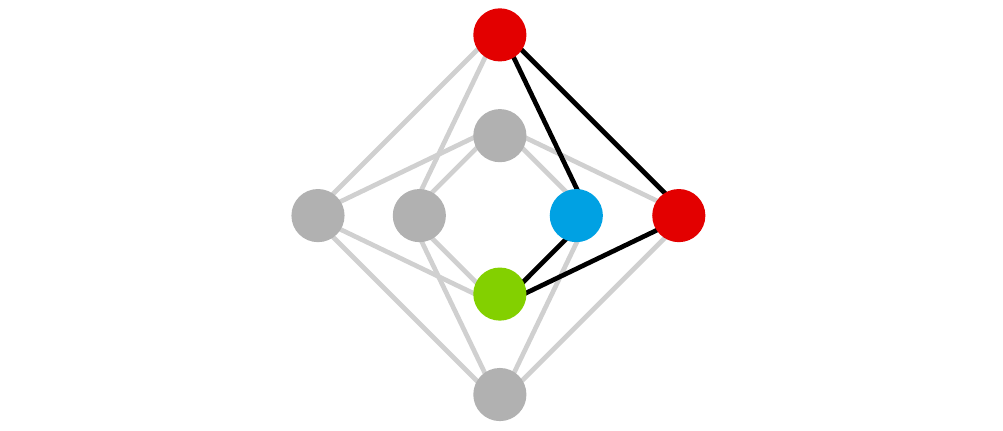}
 \caption{A graph with triangular connectivity (left) can be mapped onto the Chimera hardware graph (right) by using two physical qubits for one of the logical qubits.}
 \label{fig_triangular_embedding}
\end{figure}

\index{embedding}
As soon as problem sizes become so large that we cannot immediately find a mapping of the problem graph onto the hardware graph, it is convenient to call a dedicated routine to find this mapping for us.
Generating the optimal embedding of a graph onto another one is in general NP-hard, but a heuristic algorithm is provided by the \texttt{Ocean-SDK}. This algorithm uses probabilistic methods, which means that each time we call the function, it may return a different embedding and these may be of different quality. Thus, one typically tries several different embeddings and chooses the best one (see also \cite{calaza2021gardenoptimization,Willsch2021BenchmarkAdvantage}).
It may also be possible that the problem graph requires connectivities which are not present on the hardware graph. For instance, a triangular connectivity as shown in Fig.~\ref{fig_triangular_connectivity} cannot be mapped onto the Chimera graph.
In such a case, we have to use more than a single physical qubit (the ones on the hardware) to represent a logical qubit\index{logical and physical qubits} (the ones in the problem specification), see Fig.~\ref{fig_triangular_embedding}.
If we have more than a single physical qubit representing a logical qubit, i.e., a chain of two or more physical qubits represents a single logical qubit, the physical qubits should behave as a single entity, i.e., at the end of the annealing process, they should all have the same value.

To achieve this, the couplings $J_{ij}$ between these physical qubits are set to a negative value with large magnitude.
The magnitude (also called chain strength\index{chain strength}) determines how strongly these qubits couple and how easily the chain may ``break". In this context, a ``chain break'' means that different qubits of a chain representing the same logical qubit end up in different states.
The Ocean package includes post-processing procedures which, in this case, determine the value to return for the logical qubit by majority vote of the physical qubits. Thus, the values returned will always be valid for the original problem, although they could be far from optimal.
Too many chain breaks should be avoided as the returned solutions become ``randomized" and may be rather poor.

The optimal value of the chain strength depends on the particular problem and possibly also on the particular embedding.
If the chain strength is chosen too weak, the optimal and close-to-optimal solutions may likely contain chain breaks as it is energetically favourable to break these couplings instead of the ones that encode the actual problem instance.
In theory, the chain strength should be chosen as large as possible to satisfy the constraints.
In practice however, we have to keep in mind that the coupling strength $J_{ij}$ can only be set to values in the interval $[J_\mathrm{min},J_\mathrm{max}]$ with a limited precision (cf.~Section~\ref{sec:specification}).
Thus, all $h_i$ and $J_{ij}$ are rescaled by the factor $r$ (see Eq.~(\ref{eq:rescale})).
If the chain strength is chosen too strong, all parameters defining the problem instance will be rescaled to small values which might be no longer resolvable with the given precision.
Moreover, the energy differences of the original problem are also rescaled and become very small which could more likely lead to excitations to higher energy states. This, in turn, would lead to worse results for the original problem than if a weaker chain strength had been chosen.
The chain strength is a parameter called \texttt{chain\_strength} that can be given to the \texttt{sample\_qubo} or \texttt{sample\_ising} functions.

The \texttt{Ocean} package provides several \texttt{EmbeddingComposite} classes~\cite{dwave_documentation} to handle the generation of embeddings:
\begin{itemize}
    \item \texttt{EmbeddingComposite} tries to find an embedding each time one of the \texttt{sample} functions (\texttt{sample\_qubo} or \texttt{sample\_ising}) is called. This can be useful when submitting different problems or when studying the dependence on different embeddings.
    \item \texttt{LazyFixedEmbeddingComposite} tries to find an embedding the first time one of the \texttt{sample} functions is called. Later, it reuses this embedding. This can be useful when submitting the same problem with different hyperparameters such as annealing time or chain strength.
    \item \texttt{FixedEmbeddingComposite} takes an embedding as argument which is then used each time one of the \texttt{sample} functions is called. This can be useful when we already have an embedding for a particular problem (either by generating it ourselves or by loading a previously stored embedding) and we want to reuse it again.
    \item \texttt{TilingComposite} can be given as an argument to any of the \texttt{EmbeddingComposite}s to place several copies of small embedded graphs.
\end{itemize}

\subsection{Example: Garden optimization}

As an example, we consider a simplified four-qubit problem from the garden optimization problem presented in \cite{calaza2021gardenoptimization}. The task is to place four plants in two pots such that good neighbors are in the same pot and bad neighbors are not. The relationships between the four plants that we consider in this example are shown in Table~\ref{tab:vegetables}.

\begin{table}[t]
    \centering
    \begin{tabular}{c|c c c c}
         & Leek & Celery & Peas & Corn \\
         \hline
      Leek   &  $\circ$ & $\heartsuit$ & $\times$ & $\circ$ \\
      Celery & $\heartsuit$ & $\circ$  & $\circ$  & $\times$ \\
      Peas   & $\times$ & $\circ$ & $\circ$ & $\heartsuit$ \\
      Corn   & $\circ$ & $\times$ & $\heartsuit$ & $\circ$
    \end{tabular}
    \caption{Companion planting example of good ($\heartsuit$), neutral ($\circ$), and bad ($\times$) neighbors.}
    \label{tab:vegetables}
\end{table}

The two pots have the labels $-1$ and $+1$ and the value $s_i$ of qubit $i$ denotes into which pot we place the plant of type $i$ ($i\in \{\text{leek, celery, peas, corn}\}$).
Since we consider minimization problems, we want to minimize the energy when good neighbors are placed in the same pot.
Assume we place plants $i$ and $j$ in the same pot (i.e., $s_i=s_j$).
According to Eq.~(\ref{eq:opt_ising}), the energy is lowered if we choose $J_{ij}$ negative (which we want for good neighbors $i$ and $j$), and the energy is increased if we choose $J_{ij}$ positive (which we want for bad neighbors $i$ and $j$).
Thus, we set $J_\mathrm{leek,celery} = J_\mathrm{peas,corn}=-1$ and $J_\mathrm{leek,peas} = J_\mathrm{celery,corn} = 1$. All other $J_{ij}$ are set to zero (neutral relationship).
The energy function is then given by
\begin{align}
  E(s_\mathrm{leek}, s_\mathrm{celery}, s_\mathrm{peas}, s_\mathrm{corn}) = s_\mathrm{leek}s_\mathrm{peas} + s_\mathrm{celery}s_\mathrm{corn} - s_\mathrm{leek}s_\mathrm{celery} -s_\mathrm{peas}s_\mathrm{corn}.
\end{align}

\begin{figure}[t!]
 \centering
 \includegraphics[width=0.25\textwidth]{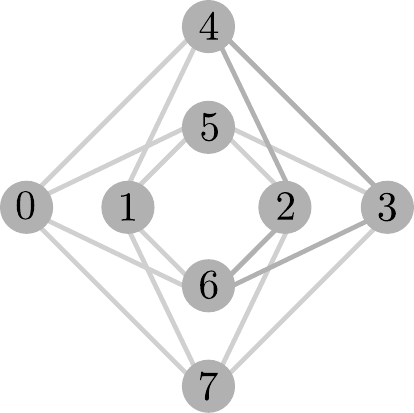}
 \caption{Labeling of the qubits in a Chimera graph unit cell of the D-Wave 2000Q quantum processors.}
 \label{fig:graph_example}
\end{figure}

The next step is to map the qubits onto the hardware graph. The labeling of the qubits in the first unit cell of the Chimera topology used in the D-Wave 2000Q processors is shown in Fig.~\ref{fig:graph_example}.
Since the problem is small and can be directly mapped onto the hardware graph, no embedding is necessary, so we define the mapping without an \texttt{EmbeddingComposite}.
We use the labeling
\begin{align}
    \text{leek} &\to 0, \quad \text{corn} \to 3, \\
    \text{celery} &\to 4, \quad \text{peas} \to 7,
\end{align}
which gives
\begin{align}
  J_{04}=J_{37} = -1 \quad\text{and}\quad J_{07} = J_{34} = 1.
\end{align}
The program to submit and solve the problem on the D-Wave 2000Q chip \texttt{DW\_2000Q\_6} is shown in Listing~\ref{example_garden}.
\\~\\
\textbf{Exercise 7:} Consider the case that leek has already been placed in the pot with label ``$-1$". Replanting it would require additional work. How could this additional cost be considered in the energy function? Modify the program in Listing~\ref{example_garden} accordingly. How does the result change?

\begin{lstlisting}[float, language=Python, caption=Example code solving the four-qubit garden problem., label=example_garden]
from dwave.system import DWaveSampler
import dwave.inspector

sampler = DWaveSampler(solver='DW_2000Q_6', token='insert_your_token_here')

h = {}

# We choose:
# 0 = leek
# 4 = celery
# 7 = peas
# 3 = corn
J = { (0,4): -1, (0,7): +1, (3,4): +1, (3,7): -1 }

response = sampler.sample_ising(h, J, num_reads=100)

dwave.inspector.show(response)
\end{lstlisting}

\newpage
\appendix
\section*{Appendix: JUQCS standard gate set}
\addcontentsline{toc}{section}{Appendix: JUQCS standard gate set}
\label{app:JUQCS}
This appendix contains the standard gate set implemented by the J\"ulich Universal Quantum Computer Simulator\index{JUQCS} (JUQCS) \cite{RAED19a}. JUQCS is a large-scale simulator for gate-based quantum computers that was also used for Google's quantum supremacy experiment \cite{Google2019QuantumSupremacy}. A GPU-accelerated version was recently used to benchmark JUWELS Booster \cite{Willsch2021JUQCSGQAOA} with over 2048 GPUs across 512 compute nodes.
A \texttt{qiskit} \cite{Qiskit} interface to JUQCS including the conversion from the \texttt{qiskit} gate set to the JUQCS gate set is available through the J\"ulich UNified Infrastructure for Quantum computing (JUNIQ) service at \url{https://jugit.fz-juelich.de/qip/juniq-platform}. This interface was used for the example programs in this lecture.

\medskip
\small

\noindent{\bf I gate}

\begin{tabular}{l  l | c}
 {\bf Description} & performs an identity operation on qubit $n$. & \multirow{ 2}{*}{$I=\left(\begin{array}{rr} 1 & 0\\ 0 & 1 \end{array}\right)$}\\
 {\bf Syntax} & I $n$ & \\
 {\bf Qiskit syntax} & \texttt{circuit.id(n)} & \\
 {\bf Argument} & $n$ integer, $0\le n < N$ with $N$ the number of qubits.& $\Qcircuit @C=1.3em @R=.4em {&\gate{I}&\qw\\}$
\end{tabular}
\medskip

\noindent{\bf H gate}

\begin{tabular}{l  l | c}
 {\bf Description} & performs a Hadamard operation on qubit $n$. & \multirow{ 2}{*}{$H=\frac{1}{\sqrt{2}}\left(\begin{array}{rr} 1 & 1\\ 1 & -1 \end{array}\right)$} \\
 {\bf Syntax} & H $n$ & \\
 {\bf Qiskit syntax} & \texttt{circuit.h(n)} & \\
 {\bf Argument} & $n$ integer, $0\le n < N$ with $N$ the number of qubits. & $\Qcircuit @C=1.3em @R=.4em {&\gate{H}&\qw\\}$
\end{tabular}
\medskip

\noindent{\bf X gate}

\begin{tabular}{l  l | c}
 {\bf Description} & performs a bit flip operation on qubit $n$. & \multirow{ 2}{*}{$X=\left(\begin{array}{rr} 0 & 1\\ 1 & 0 \end{array}\right)$} \\
 {\bf Syntax} & X $n$ & \\
 {\bf Qiskit syntax} & \texttt{circuit.x(n)} & \\
 {\bf Argument} & $n$ integer, $0\le n < N$ with $N$ the number of qubits. & $\Qcircuit @C=1.3em @R=.4em {&\gate{X}&\qw\\}$
\end{tabular}
\medskip

\clearpage
\noindent{\bf Y gate}

\begin{tabular}{l  l | c}
 {\bf Description} & performs a bit and phase flip operation on qubit $n$. & \multirow{ 2}{*}{$Y=\left(\begin{array}{rr} 0 & -i\\ i & 0 \end{array}\right)$} \\
 {\bf Syntax} & Y $n$ & \\
 {\bf Qiskit syntax} & \texttt{circuit.y(n)} & \\
 {\bf Argument} & $n$ integer, $0\le n < N$ with $N$ the number of qubits. & $\Qcircuit @C=1.3em @R=.4em {&\gate{Y}&\qw\\}$
\end{tabular}
\medskip

\noindent{\bf Z gate}

\begin{tabular}{l  l | c}
 {\bf Description} & performs a phase flip operation on qubit $n$. & \multirow{ 2}{*}{$Z=\left(\begin{array}{rr} 1 & 0\\ 0 & -1 \end{array}\right)$} \\
 {\bf Syntax} & Z $n$ & \\
 {\bf Qiskit syntax} & \texttt{circuit.z(n)} & \\
 {\bf Argument} & $n$ integer, $0\le n < N$ with $N$ the number of qubits. & $\Qcircuit @C=1.3em @R=.4em {&\gate{Z}&\qw\\}$
\end{tabular}
\medskip

\noindent{\bf S gate}

\begin{tabular}{l  l | c}
 {\bf Description} & rotates qubit $n$ about the $z$-axis by $\pi/2$. &  \multirow{ 2}{*}{$S=\left(\begin{array}{rr} 1 & 0\\ 0 & i \end{array}\right)$}\\
 {\bf Syntax} & S $n$ & \\
 {\bf Qiskit syntax} & \texttt{circuit.s(n)} & \\
 {\bf Argument} & $n$ integer, $0\le n < N$ with $N$ the number of qubits. & $\Qcircuit @C=1.3em @R=.4em {&\gate{S}&\qw\\}$
\end{tabular}
\medskip

\noindent{\bf S$^\dagger$ gate}

\begin{tabular}{l  l | c}
 {\bf Description} & rotates qubit $n$ about the $z$-axis by $-\pi/2$ & \multirow{ 2}{*}{$S^\dagger=\left(\begin{array}{rr} 1 & 0\\ 0 & -i \end{array}\right)$}\\
 {\bf Syntax} & S+ $n$ &  \\
 {\bf Qiskit syntax} & \texttt{circuit.sdg(n)} & \\
 {\bf Argument} & $n$ integer, $0\le n < N$ with $N$ the number of qubits. & $\Qcircuit @C=1.3em @R=.4em {&\gate{S^\dagger}&\qw\\}$
\end{tabular}
\medskip

\noindent{\bf T gate}

\begin{tabular}{l  l | c}
 {\bf Description} & rotates qubit $n$ about the $z$-axis by $\pi/4$ & \multirow{ 2}{*}{$T=\left(\begin{array}{cc} 1 & 0\\ 0 & (1+i)/\sqrt{2} \end{array}\right)$} \\
 {\bf Syntax} & T $n$ & \\
 {\bf Qiskit syntax} & \texttt{circuit.t(n)} & \\
 {\bf Argument} & $n$ integer, $0\le n < N$ with $N$ the number of qubits. & $\Qcircuit @C=1.3em @R=.4em {&\gate{T}&\qw\\}$
\end{tabular}
\medskip

\noindent{\bf T$^\dagger$ gate}

\begin{tabular}{l  l | c}
 {\bf Description} & rotates qubit $n$ about the $z$-axis by $-\pi/4$ & \multirow{ 2}{*}{$T^\dagger=\left(\begin{array}{cc} 1 & 0\\ 0 & (1-i)/\sqrt{2}\end{array}\right)$} \\
 {\bf Syntax} & T+ $n$ & \\
 {\bf Qiskit syntax} & \texttt{circuit.tdg(n)} & \\
 {\bf Argument} & $n$ integer, $0\le n < N$ with $N$ the number of qubits. & $\Qcircuit @C=1.3em @R=.4em {&\gate{T^\dagger}&\qw\\}$
\end{tabular}
\medskip

\noindent{\bf U1 gate}

\begin{tabular}{l  l | c}
 {\bf Description} & performs a U1($\lambda$) operation~\cite{CROS17} on qubit $n$. & \multirow{ 2}{*}{$U1(\lambda)=\left(\begin{array}{cc} 1 & 0\\ 0 & e^{i\lambda} \end{array}\!\right)$} \\
 {\bf Syntax} & U1 $n$ $\lambda$ & \\
 {\bf Qiskit syntax} & \texttt{circuit.p(lam, n)} & \\
 {\bf Arguments} & $n$ integer, $0\le n < N$ with $N$ the number of qubits, & \multirow{2}{*}{$\Qcircuit @C=1.3em @R=.4em {&\gate{U1(\lambda)}&\qw\\}$}\\
 & $\lambda$ angle in radians (floating point or integer). &
\end{tabular}
\medskip

\noindent{\bf U2 gate}

\begin{tabular}{l  l | c}
 {\bf Description} & performs a U2($\phi,\lambda$) operation~\cite{CROS17} on qubit $n$. & \multirow{ 2}{*}{$U2(\phi,\lambda)=\frac{1}{\sqrt{2}}\!
\left(\!\!\begin{array}{cc} 1 & \!-e^{i\lambda}\\ e^{i\phi} & \!e^{i(\phi+\lambda)} \end{array}\!\!\!\right)$} \\
 {\bf Syntax} & U2 $n$ $\phi$ $\lambda$ & \\
 {\bf Qiskit syntax} & \texttt{circuit.u(pi/2, phi, lam, n)} & \\
 {\bf Arguments} & $n$ integer, $0\le n < N$ with $N$ the number of qubits, & $\Qcircuit @C=1.3em @R=.4em {&\gate{U2(\phi,\lambda)}&\qw\\}$\\
 & $\phi$, $\lambda$ angles in radians (floating point or integer). &
\end{tabular}
\medskip

\clearpage
\noindent{\bf U3 gate}

\begin{tabular}{l  l | c}
 {\bf Description} & performs a U3($\theta,\phi,\lambda$) operation~\cite{CROS17} on qubit $n$. & $U3(\theta,\phi,\lambda)=$\\
 {\bf Syntax} & U3 $n$ $\theta$ $\phi$ $\lambda$ & \multirow{ 3}{*}{
$\!\!\left(\!\!\begin{array}{cc} \cos(\frac{\theta}{2}) &  -e^{i\lambda}\sin(\frac{\theta}{2})\\ e^{i\phi}\sin(\frac{\theta}{2})& e^{i(\phi+\lambda)}\cos(\frac{\theta}{2}) \end{array}\!\!\right)$} \\
 {\bf Qiskit syntax} & \texttt{circuit.u(theta, phi, lam, n)} & \\
 {\bf Arguments} & $n$ integer, $0\le n < N$ with $N$ the number of qubits, & \\
 & $\theta$, $\phi$, $\lambda$ angles in radians (floating point or integer). & $\Qcircuit @C=1.3em @R=.4em {&\gate{U3(\theta,\phi,\lambda)}&\qw\\}$
\end{tabular}
\medskip

\noindent{\bf +X gate}

\begin{tabular}{l  l | c}
 {\bf Description} & rotates qubit $n$ by $-\pi/2$ about the $x$-axis. & \multirow{ 2}{*}{$+X=\frac{1}{\sqrt{2}}\left(\begin{array}{rr} 1 & i\\ i & 1 \end{array}\right)$} \\
 {\bf Syntax} & +X $n$ & \\
 {\bf Qiskit syntax} & \texttt{circuit.rx(-pi/2, n)} & \\
 {\bf Argument} & $n$ integer, $0\le n < N$ with $N$ the number of qubits. & $\Qcircuit @C=1.3em @R=.4em {&\gate{+X}&\qw\\}$
\end{tabular}
\medskip

\noindent{\bf -X gate}

\begin{tabular}{l  l | c}
 {\bf Description} & rotates qubit $n$ by $+\pi/2$ about the $x$-axis. & \multirow{ 2}{*}{$-X=\frac{1}{\sqrt{2}}\left(\begin{array}{rr} 1 & -i\\ -i & 1 \end{array}\right)$} \\
 {\bf Syntax} & -X $n$ & \\
 {\bf Qiskit syntax} & \texttt{circuit.rx(pi/2, n)} & \\
 {\bf Argument} & $n$ integer, $0\le n < N$ with $N$ the number of qubits. & $\Qcircuit @C=1.3em @R=.4em {&\gate{-X}&\qw\\}$
\end{tabular}
\medskip

\noindent{\bf +Y gate}

\begin{tabular}{l  l | c}
 {\bf Description} & rotates qubit $n$ by $-\pi/2$ about the $y$-axis. & \multirow{ 2}{*}{$+Y=\frac{1}{\sqrt{2}}\left(\begin{array}{rr} 1 & 1\\ -1 & 1 \end{array}\right)$} \\
 {\bf Syntax} & +Y $n$ & \\
 {\bf Qiskit syntax} & \texttt{circuit.ry(-pi/2, n)} & \\
 {\bf Argument} & $n$ integer, $0\le n < N$ with $N$ the number of qubits. & $\Qcircuit @C=1.3em @R=.4em {&\gate{+Y}&\qw\\}$
\end{tabular}
\medskip

\noindent{\bf -Y gate}

\begin{tabular}{l  l | c}
 {\bf Description} & rotates qubit $n$ by $+\pi/2$ about the $y$-axis. &  \multirow{ 2}{*}{$-Y=\frac{1}{\sqrt{2}}\left(\begin{array}{rr} 1 & -1\\ 1 & 1 \end{array}\right)$} \\
 {\bf Syntax} & -Y $n$ & \\
 {\bf Qiskit syntax} & \texttt{circuit.ry(pi/2, n)} & \\
 {\bf Argument} & $n$ integer, $0\le n < N$ with $N$ the number of qubits. & $\Qcircuit @C=1.3em @R=.4em {&\gate{-Y}&\qw\\}$
\end{tabular}
\medskip

\noindent{\bf R(k) gate}

\begin{tabular}{l  l | c}
 {\bf Description} & changes the phase of qubit $n$ by $2\pi/2^k$. & \multirow{ 2}{*}{$R(k)=\left(\begin{array}{cc} 1 & 0\\ 0 & e^{2\pi i/2^k} \end{array}\!\right)$} \\
 {\bf Syntax} & R $n$ $k$ & \\
 {\bf Qiskit syntax} & \texttt{circuit.p(2*pi/2**k, n)} & \\
 {\bf Arguments} & $n$ integer, $0\le n < N$ with $N$ the number of qubits, & \multirow{ 2}{*}{$\Qcircuit @C=1.3em @R=.4em {&\gate{R(k)}&\qw\\}$} \\
 & $k$ is non-negative integer. &
\end{tabular}
\medskip

\noindent{\bf R$^\dagger$(k) gate}

\begin{tabular}{l  l | c}
 {\bf Description} & changes the phase of qubit $n$ by $-2\pi/2^k$. & \multirow{ 2}{*}{$R^\dagger(k)=\left(\begin{array}{cc} 1 & 0\\ 0 & e^{-2\pi i/2^k} \end{array}\!\right)$} \\
 {\bf Syntax} & R $n$ $-k$ & \\
 {\bf Qiskit syntax} & \texttt{circuit.p(-2*pi/2**k, n)} & \\
 {\bf Arguments} & $n$ integer, $0\le n < N$ with $N$ the number of qubits, & \multirow{ 2}{*}{$\Qcircuit @C=1.3em @R=.4em {&\gate{R^\dagger(k)}&\qw\\}$} \\
 & $k$ is non-negative integer. &
\end{tabular}
\medskip

\clearpage
\noindent{\bf CNOT gate}

\begin{tabular}{l  l | c}
 {\bf Description} & flips the target qubit if the control qubit is 1. & $\mathrm{CNOT}=$ \\
 {\bf Syntax} & CNOT $control$ $target$ & \multirow{ 4}{*}{$\left(\begin{array}{cccc} 1 & 0 & 0 & 0\\ 0 & 1 & 0 & 0 \\ 0 & 0 & 0 & 1\\0 & 0 & 1 & 0\end{array}\right)$}\\
 {\bf Qiskit syntax} & \texttt{circuit.cx(control, target)} & \\
 {\bf Arguments} & $control\not=target$ integers in the range & \\
 & $0,\dots,N-1$ with $N$ the number of qubits. &  \\
 {\bf Note} & The matrix looks different from the \texttt{qiskit} documen- & \\
 & tation as \texttt{qiskit} uses the ordering $\ket{q_{n-1}\cdots q_{0}}$ while & \multirow{2}{*}{$\Qcircuit @C=1.3em @R=.4em {&\ctrl{1}&\qw\\&\targ&\qw}$}\\
 & all standard text books as well as these lecture notes use & \\
 & $\ket{q_0\cdots q_{n-1}}$. To fix this, the supplied example programs & \\
 & invoke the \texttt{circuit.reverse\_bits()} function.
\end{tabular}
\medskip

\noindent{\bf U(k) gate}

\begin{tabular}{l  l | c}
 {\bf Description} & shifts the phase of the target qubit by $2\pi/2^k$ if the & $U(k)=$\\
 & control qubit is 1. &  \multirow{ 4}{*}{$\left(\begin{array}{cccc}
1 & 0 & 0 & 0\\
0 & 1 & 0 & 0\\
0 & 0 & 1 & 0\\
0 & 0 & 0 & e^{2\pi i/2^k}
\end{array}\!\!\right)$}\\
 {\bf Syntax} & U $control$ $target$ $k$ & \\
 {\bf Qiskit syntax} & \texttt{circuit.cp(2*pi/2**k, control, target)} & \\
 {\bf Arguments} & $control\not=target$ integers in the range $0,\dots,N-1$ &  \\
 & with $N$ the number of qubits, $k$ non-negative integer. &\multirow{2}{*}{ $\Qcircuit @C=1.3em @R=.4em {&\ctrl{1}&\qw\\&\gate{U(k)}&\qw}$}\\
 &
\end{tabular}
\medskip

\noindent{\bf U$^\dagger$(k) gate}

\begin{tabular}{l  l | c}
 {\bf Description} & shifts the phase of the target qubit by $-2\pi/2^k$ if the &  $U^\dagger(k)=$\\
 & control qubit is 1. & \multirow{ 4}{*}{$\left(\begin{array}{cccc}
1 & 0 & 0 & 0\\
0 & 1 & 0 & 0\\
0 & 0 & 1 & 0\\
0 & 0 & 0 & e^{-2\pi i/2^k}
\end{array}\!\!\!\right)$}\\
 {\bf Syntax} & U $control$ $target$ $-k$ & \\
 {\bf Qiskit syntax} & \texttt{circuit.cp(-2*pi/2**k, control,target)} & \\
 {\bf Arguments} & $control\not=target$ integers in the range $0,\dots,N-1$ & \\
 & with $N$ the number of qubits, $k$ non-negative integer.& \multirow{2}{*}{$\Qcircuit @C=1.3em @R=.4em {&\ctrl{1}&\qw\\&\gate{U^\dagger(k)}&\qw}$} \\
 &
\end{tabular}
\medskip

\noindent{\bf Toffoli gate}

\begin{tabular}{l  l | c}
 {\bf Description} & flips the target qubit if both control qubits are 1. & $\mathrm{TOFFOLI}=$\\
 {\bf Syntax} & TOFFOLI $control_1$ $control_2$ $target$ & \multirow{8}{*}{$\!\left(\!\begin{array}{cccccccc}
1 & \!0 & \!0 & \!0& \!0 & \!0 & \!0& \!0 \\
0 & \!1 & \!0 & \!0& \!0 & \!0 & \!0& \!0 \\
0 & \!0 & \!1 & \!0& \!0 & \!0 & \!0& \!0 \\
0 & \!0 & \!0 & \!1& \!0 & \!0 & \!0& \!0 \\
0 & \!0 & \!0 & \!0& \!1 & \!0 & \!0& \!0 \\
0 & \!0 & \!0 & \!0& \!0 & \!1 & \!0& \!0 \\
0 & \!0 & \!0 & \!0& \!0 & \!0 & \!0& \!1 \\
0 & \!0 & \!0 & \!0& \!0 & \!0 & \!1& \!0 \\
\end{array} \!\right)$} \\
 {\bf Qiskit syntax} & \texttt{circuit.ccx(control1,control2,target)} & \\
 {\bf Arguments} & $control_1\not=control_2\not=target\not=control_1$ integers in & \\
 & the range $0,\dots,N-1$ with $N$ the number of qubits. & \\
 {\bf Note} & The matrix looks different from the \texttt{qiskit} & \\
 & documentation because \texttt{qiskit} uses the ordering & \\
 & $\ket{q_{n-1}\cdots q_{0}}$ while all standard text books as well as & \\
 & these lecture notes use $\ket{q_0\cdots q_{n-1}}$. To fix this, the &\\
 & supplied example programs invoke the &  \multirow{2}{*}{$\Qcircuit @C=1.3em @R=.4em {&\ctrl{1}&\qw\\ &\ctrl{1}&\qw\\ &\targ&\qw}$}\\
 & \texttt{circuit.reverse\_bits()} function.
\end{tabular}
\medskip
\normalsize

\section*{Acknowledgments}
\addcontentsline{toc}{section}{Acknowledgments}
We would like to thank Carlos D.~Gonzalez Calaza, Nils K\"uchler, and Cameron Perot for helpful comments and suggestions on these lecture notes.

\clearpage
\printbibliography[heading=bibintoc]

\end{document}

%% file: custom_bib.tex
\DeclareBibliographyDriver{article}{%
  \usebibmacro{bibindex}%
  \usebibmacro{begentry}%
  \usebibmacro{author/translator+others}%
  \setunit{\addcomma\space}\newblock%
  \usebibmacro{title}
  \setunit{\addcomma\space}\newblock%
  \usebibmacro{journal}%
  \setunit*{\addspace}%
  \printfield{volume}%
  \setunit{\addcomma\linebreak[1]\space}%
  \printfield{pages}%
  \setunit*{\addspace}%
  \usebibmacro{articledate}%
  \setunit{\bibpagerefpunct}\newblock%
  \usebibmacro{pageref}%
  \newunit\newblock%
  \usebibmacro{finentry}%
}
\DeclareBibliographyDriver{unpublished}{%
  \usebibmacro{bibindex}%
  \usebibmacro{begentry}%
  \usebibmacro{author}%
  \setunit{\addcomma\space}\newblock%
  \usebibmacro{title}
  \setunit{\addcomma\space}\newblock%
  \printtext{\thefield{eprint}}%
  \setunit*{\addspace}%
  \usebibmacro{articledate}
  \setunit{\bibpagerefpunct}\newblock%
  \usebibmacro{pageref}%
  \newunit\newblock%
  \usebibmacro{finentry}%
}
\DeclareBibliographyDriver{online}{%
  \usebibmacro{bibindex}%
  \usebibmacro{begentry}%
  \usebibmacro{author}%
  \setunit{\addcomma\space}\newblock%
  \usebibmacro{title}
  \setunit{\addcomma\space}\newblock%
  \url{\thefield{url}}%
  \setunit*{\addcomma\addspace}%
  \printtext{\thefield{note}}
  \setunit{\bibpagerefpunct}\newblock%
  \usebibmacro{pageref}%
  \newunit\newblock%
  \usebibmacro{finentry}%
}
\DeclareBibliographyDriver{book}{%
  \usebibmacro{bibindex}%
  \usebibmacro{begentry}%
  \usebibmacro{author/editor+others/translator+others}%
  \setunit{\addcomma\space}\newblock%
  \usebibmacro{maintitle+title}
  \setunit{\addcomma\space}\newblock%
  \printlist{publisher}%
  \setunit*{\addcomma\space}%
  \usebibmacro{date}%
  \setunit{\bibpagerefpunct}\newblock%
  \usebibmacro{pageref}%
  \newunit\newblock%
  \usebibmacro{finentry}%
}
\DeclareBibliographyDriver{inproceedings}{%
  \usebibmacro{bibindex}%
  \usebibmacro{begentry}%
  \usebibmacro{author/translator+others}%
  \setunit{\addcomma\space}\newblock%
  \usebibmacro{title}%
  \setunit{\addcomma\space}\newblock%
  \printfield{booktitle}%
  \iffieldundef{series}%
    {%
      \iffieldundef{volume}%
      {}%
      {\setunit{\addcomma\space}%
      \printfield{volume}}%
    }%
    {%
      \setunit{\addcomma\space}%
      \printfield{series}%
      \iffieldundef{volume}%
      {}%
      {%
        \setunit*{\addspace}%
        \printfield{volume}%
      }%
    }%
  \iffieldundef{pages}
    {}%
    {\setunit{\addcomma\space}%
    \printtext{pp.}
    \printfield{pages}}%
  \ifnameundef{editor}
    {}%
    {\setunit{\addcomma\space}%
    \usebibmacro{byeditor+others}}%
  \setunit*{\addspace}%
  \usebibmacro{articledate}%
  \setunit{\bibpagerefpunct}\newblock%
  \usebibmacro{pageref}%
  \newunit\newblock%
  \usebibmacro{finentry}%
}
\DeclareBibliographyDriver{thesis}{%
  \usebibmacro{bibindex}%
  \usebibmacro{begentry}%
  \usebibmacro{author}%
  \setunit{\printdelim{nametitledelim}}\newblock
  \usebibmacro{title}%
  \newunit\newblock
  \printfield{type}%
  \newunit
  \usebibmacro{institution+location+date}%
  \newunit\newblock
  \usebibmacro{addendum+pubstate}%
  \setunit{\bibpagerefpunct}\newblock
  \usebibmacro{pageref}%
    \newunit\newblock%
  \usebibmacro{finentry}
}
\newbibmacro{string+doiurlisbn}[1]{%
  \iffieldundef{doi}{%
    \iffieldundef{url}{%
      \iffieldundef{isbn}{%
        \iffieldundef{issn}{%
          \iffieldundef{eprint}{%
           {#1}%
            }{\href{https://arxiv.org/abs/\thefield{eprint}}{#1}}%
          }{\href{https://books.google.com/books?vid=ISSN\thefield{issn}}{#1}}%
      }{\href{https://books.google.com/books?vid=ISBN\thefield{isbn}}{#1}}%
    }{\href{\thefield{url}}{#1}}%
  }{\href{https://dx.doi.org/\thefield{doi}}{#1}}%
}
\DeclareFieldFormat{linked}{\usebibmacro{string+doiurlisbn}{#1}}
\renewbibmacro*{date}{\iffieldundef{year}{}{\printtext[linked]{\printdate}}}
\newbibmacro*{articledate}{\printtext[parens]{\printtext[linked]{\printdate}}}

\renewcommand{\bibpagerefpunct}{\addspace}
\renewbibmacro*{pageref}{%
  \iflistundef{pageref}
    {}
    {\printtext[brackets]{
       \ifnumgreater{\value{pageref}}{1}
         {\bibstring{backrefpages}\ppspace}
         {\bibstring{backrefpage}\ppspace}%
       \printlist[pageref][-\value{listtotal}]{pageref}}}
}
\renewbibmacro{in:}{} 
\DefineBibliographyStrings{english}{ 
  page             = {},
}
\DeclareFieldFormat{pages}{\mkfirstpage[{\mkpageprefix[bookpagination]}]{#1}} 
\DeclareNolabel{ 
  \nolabel{\regexp{[\p{Z}\p{P}\p{S}\p{C}]+}}
}
\renewbibmacro*{author}{%
  \iffieldundef{usera}{%
    \textsc{\printnames{author}}%
  }{%
    \printnames{author} (\printfield{usera})%
  }
}
\DeclareFieldFormat[inproceedings]{volume}{\mkbibemph{Vol.~#1}}
\DeclareFieldFormat[inproceedings]{series}{\mkbibemph{#1}}
\DeclareFieldFormat[article,inproceedings,book]{volume}{\textbf{#1}}
\DeclareFieldFormat[article,inproceedings,unpublished,book,thesis]{title}{\mkbibemph{#1}}
\DeclareFieldFormat[article,inproceedings,book]{journaltitle}{#1}
\DeclareFieldFormat[article,inproceedings,book]{booktitle}{#1}
\DeclareFieldFormat[article,inproceedings,book,unpublished,thesis]{author}{\textsc{#1}}